\documentclass[10pt]{article}\usepackage[margin=1.2in]{geometry}
\usepackage{amssymb}
\usepackage{theorem}
\usepackage{verbatim}
\usepackage{amsmath}
\usepackage{tikz}
\newtheorem{theorem}{Theorem}
\newtheorem{definition}[theorem]{Definition}
\newtheorem{lemma}[theorem]{Lemma}

\newtheorem{proposition}[theorem]{Proposition}

\newtheorem{open-probl}{Open problem}

\numberwithin{equation}{section}
\newcommand{\p}{{\bf p}}    \newcommand{\q}{{\bf q}}          \newcommand{\w}{{\bf w}}    
   
\newcommand{\N}{\mathbb{N}}

\newcommand{\Q}{\mathbb{Q}}
\newcommand{\Z}{\mathbb{Z}}
\newcommand{\R}{\mathbb{R}}
\newcommand{\C}{\mathbb{C}}

\newcommand{\card}{\mathop{\mathrm{Card}}\nolimits}

\newtheorem{exampleA}[theorem]{Example}
\newtheorem{remarkA}[theorem]{Remark}
\newenvironment{example}%
{\begin{exampleA}\begin{normalfont}}{\end{normalfont}\end{exampleA}}
\newenvironment{remark}%
    {\begin{remarkA}\begin{normalfont}}{\end{normalfont}\end{remarkA}}
\newenvironment{proof}[1][Proof.]
    {\addvspace{\bigskipamount}\noindent\emph{#1} }%
    {\par\addvspace{\bigskipamount}}

\newcommand{\qed}{\ifmmode\raisebox{3.3pt}{\fbox{}}\else{\ifhmode%
\unskip\fi\nobreak\hfil\penalty50\hskip1em\null\nobreak\hfil\raisebox%
{3.3pt}{\fbox{}}\parfillskip=0pt\finalhyphendemerits=0\endgraf}\fi}

\begin{document}
\title{Rational decomposition of polyhedra and a conjecture of Brlek and Reutenauer}
 
 \author{
  Flavio D'Alessandro\footnote{dalessan@mat.uniroma1.it }
  \and
    Paolo Papi \footnote{papi@mat.uniroma1.it}
    \and
Dipartimento di Matematica ``G. Castelnuovo" \\
 Sapienza Universit\`a di Roma\\
  Piazzale Aldo Moro 2, 00185 Roma,  Italy
  }

\date{}
\maketitle

\begin{abstract}
\noindent
We  prove in the affirmative a conjecture of  Brlek and Reutenauer  \cite{BerReut}, which relates unambiguous rational sets  to a well
known result due to Khovanskii and Pukhlikov  \cite{Ko-Pu-1992}  and  independently to  Lawrence \cite{Law-1990}
 on the valuations of the algebra
 spanned by the indicator functions of   polyhedra.
\end{abstract}

\section{Introduction}\label{unosec}

Let $P$ be a rational polyhedron in $\R^d$, that is, a polyhedron defined by finitely many linear inequalities with rational
coefficients, and let $X_P=P\cap\Z^d$ be the set of its integer points. This paper compares two ways of associating a rational function
with $P$: an analytic one and a combinatorial one.

The analytic one is classical. Write each ${\bf m}=(m_1,\ldots,m_d)\in\Z^d$ as the Laurent monomial
${\bf x}^{\bf m}=x_1^{m_1}\cdots x_d^{m_d}$. If $P$ contains no lines, its {\em generating series} $\sum_{{\bf m}\in X_P}{\bf x}^{\bf m}$
converges absolutely on a non-empty open subset of $\C^d$ to a rational function and, by a theorem of Khovanskii and
Pukhlikov \cite{Ko-Pu-1992} and, independently, of Lawrence \cite{Law-1990}, this defines a valuation $\mathcal F$, that is, a linear
map, on the algebra ${\cal P}(\Q^d)$ spanned by the indicator functions of the rational polyhedra, which vanishes on the polyhedra
containing a line. We call $\mathcal F$ the valuation (KPL).

The combinatorial one is due to Brlek and Reutenauer \cite{BerReut}. The set $X_P$ is a rational subset of $\Z^d$: it is obtained from
finite sets by finitely many unions, sums and stars, the star of a set being the submonoid that it generates. By a theorem of Eilenberg
and Sch\"utzenberger \cite{ES}, $X_P$ is also {\em unambiguous}: it is described by a rational expression which produces each of its
elements exactly once. Reading such an expression with the rules $\cup\mapsto+$ and $T^*\mapsto(1-T)^{-1}$, one obtains a rational
function; Brlek and Reutenauer proved that this function depends only on $X_P$, and not on the expression, and that it defines a
valuation $\varphi$ on ${\cal P}(\Q^d)$, which also vanishes on the polyhedra containing a line. We call $\varphi$ the valuation (BR).
For instance, the integer points of the half-line $\{t\geq 0\}\subseteq\R$ form the set $\N$, described by the expression $x^*$, and both
valuations give $(1-x)^{-1}=\sum_{n\geq 0}x^n$; unambiguity is essential, since the ambiguous expression $x^*(x^2)^*$ of $\N$ would give
$(1-x)^{-1}(1-x^2)^{-1}$.

Brlek and Reutenauer conjectured that the valuations (BR) and (KPL) coincide \cite{BerReut}. {\em In the present paper we prove this
conjecture} (Theorem \ref{th:MAIN-B}). As a consequence, the function (KPL) of a polyhedron, an analytic object, is obtained by a purely
combinatorial reading of {\em any} unambiguous rational expression of its integer points.

The proof rests on a simple observation (Lemma \ref{lm:simple-dec}). Let $P$ contain no lines, let $K$ be its recession cone, and assume
that $X_P$ is the disjoint union of finitely many {\em simple sets}
$${\bf p}_i+\N{\bf v}_{i,1}+\cdots+\N{\bf v}_{i,r_i}, \qquad\quad i=1,\ldots,t,$$
where ${\bf p}_i\in\Z^d$ and, for each $i$, the periods ${\bf v}_{i,1},\ldots,{\bf v}_{i,r_i}$ are linearly independent vectors of
$K\cap\Z^d$. Then both $\varphi([P])$ and the sum of the generating series of $P$ are given by the same formula
$$\sum_{i=1}^t\frac{{\bf x}^{{\bf p}_i}}{(1-{\bf x}^{{\bf v}_{i,1}})\cdots(1-{\bf x}^{{\bf v}_{i,r_i}})}:$$
the decomposition gives an unambiguous rational expression of $X_P$, and the condition on the periods is exactly what makes the
generating series converge. Decompositions of this kind exist by the theorem of Eilenberg and Sch\"utzenberger, since the periods of a
linear subset of $X_P$ automatically lie in $K$ (Remark \ref{rmk:ES}). Our main technical result, Theorem \ref{cor:MainDecomp}, makes
them effective: the decomposition is computed directly from the inequalities defining $P$, by a recursive construction of depth at most
$2d$, which slices $P$ by parallel hyperplanes and partitions it into chambers, and whose pieces are translates of faces of $K$
(Proposition \ref{prop:MainDecomp} and Examples \ref{ex:dim-3-case-2} and \ref{ex:case-1}). In this way one obtains an explicit unambiguous rational
expression of $X_P$, whose (BR) reading is the function (KPL) of $P$.

As a decomposition of polyhedra, the one of Theorem \ref{cor:MainDecomp} is close to known constructions: it is a ``half-open'', i.e.
sign-free, decomposition of a pointed rational polyhedron into affine simplicial cones, for which we refer to
\cite{KoeVer,BeckSottile,Koeppe07,BrunsIchim} and, for the integer points of a simplicial cone through its fundamental parallelepiped,
to \cite{Stanley82}; an effective version, with bounds on the number of the pieces, is stated in \cite[Prop. 4.2]{NguyenPak} on the
basis of \cite{Meiser}, the same decomposition appears in \cite[Prop. 3]{dis2009}, and decompositions of this shape are also called
{\em affine stratifications} \cite{Miller13}. The point of the present paper is that such decompositions, with integral apexes and with
all the periods in the recession cone, compute the valuation (BR); we give a complete and self-contained proof of
Theorem \ref{cor:MainDecomp} in this form. A second proof of the conjecture is obtained from a theorem of Brion \cite{Brion}
(Section \ref{sec:brion}), which expresses the indicator function of $P$, up to polyhedra with lines, through the tangent cones of $P$ at
its vertices: since both valuations vanish on polyhedra with lines, one is reduced to cones with rational apex, which are treated in
Section \ref{sec:cones}. Brion's identity involves subtractions of indicator functions, according to the inclusion--exclusion principle,
whereas the first proof only uses disjoint unions of subsets of $X_P$.

{\em Background.} Unambiguity of rational sets is a classical theme of theoretical computer science. It depends on the
monoid \cite{BR88,Saka09}; it holds in free monoids, as a consequence of fundamental results of automata theory
\cite{BR88,Eil-74,Saka09}, and in finitely generated commutative monoids, by the theorem of Eilenberg and Sch\"utzenberger \cite{ES},
proved for $\N^d$, independently, by Ito \cite{ito} (see \cite{Saka09} for a thorough account). By a theorem of Ginsburg and
Spanier \cite{GS} (see also \cite{GS-book}), the rational subsets of $\Z^d$, also called semilinear sets, are exactly the subsets which
are definable by first-order formulas of Presburger arithmetic. Semilinear sets are still the object of active research: Chistikov and
Haase \cite{ChistikovHaase} set up a unified framework to bound the size of the representations of semilinear sets obtained by several
operations and decompositions, including unambiguous ones; Nguyen and Pak \cite{NguyenPak} gave an algorithm which computes an
unambiguous representation of a semilinear set $X$ in time polynomial in the size of the period length of $X$; and the theorem of
Eilenberg and Sch\"utzenberger has been applied to the exact description of vector partition functions \cite{dis2008,dis2009} and to the
approximation of semilinear sets by recognizable sets \cite{di2013-diof}. Unambiguity also plays a natural role in enumerative
combinatorics, where counting requires that every object be represented in a unique way (see \cite{DeLHemKop2013}, Sec. 2.6), and in
particular in the enumeration of the integer points of rational polyhedra, where Barvinok's algorithm counts the integer points of a
rational polytope in polynomial time when the dimension is fixed \cite{Bar1} (see also \cite{Bar-08,Bar-Pom}).

Finally, \cite{BerReut} contains a second conjecture, of analytic nature, stated right after the first one at the
end of Section 8 of that paper. There the construction recalled
above is extended from the rational subsets of $\Z^d$ to the \emph{rational series} on $\Z^d$ with coefficients in a field ${\bf F}$,
that is, to the formal sums $S=\sum_{{\bf v}\in\Z^d}\alpha_{\bf v}\,{\bf x}^{\bf v}$, $\alpha_{\bf v}\in{\bf F}$, obtained from Laurent
polynomials by sums, products and stars, whenever these operations are defined. With every such series $S$ a rational function
$f_S\in{\bf F}(x_1,\ldots,x_d)$ is associated, which depends only on $S$ (\cite{BerReut}, Theorem 5); if
$S=\sum_{{\bf m}\in X}{\bf x}^{\bf m}$ is the characteristic series of a rational subset $X$ of $\Z^d$, then $f_S$ is the function
obtained from any unambiguous rational expression of $X$. The second conjecture states that, if ${\bf F}$ is complete with respect to an
absolute value and the series $\sum_{{\bf v}\in\Z^d}\alpha_{\bf v}\,{\bf a}^{\bf v}$ converges absolutely for some
${\bf a}=(a_1,\ldots,a_d)\in{\bf F}^d$ with non-zero coordinates, then its sum is
$f_S({\bf a})$.

The two conjectures are related as follows: the first conjecture is equivalent to the second one restricted to ${\bf F}=\C$,
with its usual absolute value, and to the
characteristic series $S_P=\sum_{{\bf m}\in X_P}{\bf x}^{\bf m}$ of the sets $X_P$. Indeed, let $P$ be a rational polyhedron without
lines. Then $S_P$ converges absolutely on a non-empty open subset of $\C^d$ and, wherever it converges absolutely, its sum is the value
of the function (KPL) of $P$ (see Remark \ref{rmk:conv} and Theorem \ref{th:KPL}). Hence the second conjecture for $S_P$ says that the functions (BR) and (KPL) of
$P$ take the same values on a non-empty open set, that is, that they coincide; conversely, if they coincide, then the sum of $S_P$ is
$f_{S_P}({\bf a})$ at every point ${\bf a}$ of absolute convergence. If instead $P$ contains a line, then both valuations vanish on $P$,
while $S_P$, unless it is zero, converges absolutely at no point, so that the second conjecture gives no information. In these terms,
Lemma \ref{lm:simple-dec} below, together with Theorem \ref{cor:MainDecomp}, proves the second conjecture for the series $S_P$ on a non-empty open subset of $\C^d$, which is
what the first conjecture requires. The second conjecture in its general form, for rational series with arbitrary coefficients, is
not addressed in this paper: we plan to deal with it in a subsequent publication.

The paper is organized as follows. Section \ref{due} recalls the rational subsets of $\Z^d$, their relation with Presburger
arithmetic, and the few facts on polyhedra that we need. Section \ref{cinque} presents the valuations (BR) and (KPL) and the
conjecture. In Section \ref{MT} we prove Lemma \ref{lm:simple-dec}, settle the case of cones, and deduce the Main Theorem from
Theorem \ref{cor:MainDecomp}, whose proof, with examples in dimension $3$ and $2$, occupies Section \ref{sei:4}. Section \ref{sec:brion}
contains the proof based on Brion's identity.

   \section{Preliminaries and basic definitions}\label{due}
    We first recall the notion of rational set by following \cite{BR88,Eil-74,ES,Saka09}.

 \subsection{Rational sets of a monoid}\label{due-1}
We denote by $\N$ the set of non-negative integers. Let $M$ be a monoid. A subset of $M$ is {\em rational} if it is obtained from
finite subsets of $M$ by finitely many applications of the {\em rational operations}: finite union, product $XY=\{xy : x\in X, \ y\in Y\}$,
and star $X^*=\bigcup_{n\in\N}X^n$, the submonoid generated by $X$. The product $XY$ is {\em unambiguous} if $xy=x'y'$, with
$x,x'\in X$ and $y,y'\in Y$, implies $x=x'$ and $y=y'$. The family of {\em unambiguous rational} subsets of $M$ is the smallest family of
subsets of $M$ which contains the finite sets and is closed under disjoint finite unions, unambiguous products, and stars of the sets which are
bases of free submonoids of $M$.

\subsection{Rational sets of $\Z^d$}\label{due-1-1}
We shall  identify free abelian groups (resp. monoids) of rank $d$  with $\Z^d$ (resp. $\N^d$)  with the usual additive structure. The operation of addition is extended from elements to subsets.
Let  $B =\{{\bf v}_1, \dots , {\bf v}_n\}$ be a finite subset of $\Z^d$ (resp., $\N^d$). Set $${\bf v}^\oplus=\mathbb N \bf v.$$ Then we
denote by $B^ \oplus$ the submonoid of $\Z^d$ (resp., $\N^d$) generated by $B$, that is
 \begin{equation}\label{eq:rat-oplus-Zd}
 B^ \oplus \;=\; {\bf v}^\oplus_1 + \cdots + {\bf v}^\oplus_n \;=\; \{m_1{\bf v}_1 + \cdots + m_n {\bf v}_n
\;\mid\; m_i \in \N\}.
                \end{equation}
The following definitions are useful.
\begin{definition}\label{def-semi-simple-set}
Let $X$ be a subset of $\Z^d$ (resp., $\N^d$). Then
\begin{enumerate}
\item[(i)]  $X$ is   {\em linear} in $\Z^d$ (resp., $\N^d$) if
$$X = {\bf v_0} + \{{\bf v}_1, \dots , {\bf v}_n\}^\oplus,$$
where
${\bf v_0}, {\bf v}_1, \dots , {\bf v}_n$ are vectors of $\Z^d$,
 \item[(ii)] $X$ is {\em simple} in $\Z^d$
 if, moreover,  the vectors ${\bf v}_1, \dots , {\bf v}_n$ are linearly independent in $\Q ^d$,
\item[(iii)] $X$ is 
{\em semi-linear} in $\Z^d$  if $X$ is a finite union of linear sets in $\Z^d$,

\item[(iv)] $X$ is 
{\em semi-simple} in $\Z^d$  if $X$ is a finite disjoint union of simple sets in $\Z^d$.
\end{enumerate}
\end{definition}
  We shall use the same terminology for subsets of  $\N^d$.\par
  The following characterization of rational sets has been proven by Eilenberg and Sch\"utzenberger in   \cite{ES}  and, independently, in the case of $\N^d$, by Ito in  \cite{ito}.

 \begin{theorem}
\label{ratmGS}
Let $X$ be a subset of $\Z^d$ (resp., $\N^d$). Then
 $X$ is  semi-linear in $\Z^d$ (resp., $\N^d$) if and only if
 $X$ is  semi-simple  in $\Z^d$ (resp., $\N^d$).
 \end{theorem}
It is easily checked that a rational set in $\Z^d$ (resp., $\N^d$) is semi-linear (cf  \cite{Saka09}, Ch. II, Sec. 7, Prop. 7.15).
 Hence, by Definition \ref{def-semi-simple-set} and Theorem \ref{ratmGS}, rational, unambiguous rational,
 semi-linear and semi-simple sets are equivalent  (see \cite{dis2009} for a recent viewpoint).
 In the sequel, we may refer to such sets by using all the terms above, without any further justification.

Writing the elements ${\bf m}=(m_1,\ldots,m_d)$ of $\Z^d$ multiplicatively, as Laurent monomials
${\bf x}^{\bf m}=x_1^{m_1}\cdots x_d^{m_d}$ in the variables $x_1,\ldots,x_d$, a {\em rational expression} of a rational subset $X$ of
$\Z^d$ is, in this paper, a formula of the form
$$\bigcup _ {i=1}^t\ {\bf x}^{{\bf v}_{0i}}\,({\bf x}^{{\bf v}_{1i}})^*\cdots ({\bf x}^{{\bf v}_{n_ii}})^*$$
which describes $X$, the star being interpreted as the operator $(^\oplus)$ of Eq. (\ref{eq:rat-oplus-Zd}). The expression is
{\em unambiguous} if every element of $X$ is obtained from it in a unique way, that is, if the sets of the union are pairwise disjoint
and, for each $i$, the vectors ${\bf v}_{1i},\ldots,{\bf v}_{n_ii}$ are linearly independent. By Theorem \ref{ratmGS}, every rational
subset of $\Z^d$ has an unambiguous rational expression, which is not unique: for instance, $x^*y^*$, $\ y^* \cup xx^* \cup xyx^*y^*$
and $\ y^* \cup xy^* \cup x^2x^*y^*$ are unambiguous rational expressions of $\N^2$.
                  \subsection{Rational sets of $\Z^d$ and the Presburger Arithmetic}\label{due-3}

Denote by ${\cal Z} = \langle \Z; = ; < ; + ; 0; 1\rangle$ and by ${\cal N} = \langle \N; = ; + ; 0; 1\rangle$ the standard and the
positive {\em Presburger arithmetic}, respectively. A subset $X$ of $\Z^d$ (resp., $\N^d$) is {\em first-order definable} in $\cal Z$
(resp., $\cal N$) if $X = \{ (x_1, \dots, x_d) : \Phi(x_1, \dots, x_d)\}$ for a first-order formula $\Phi$ of $\cal Z$ (resp.,
$\cal N$) with at most $d$ free variables. By a theorem of Ginsburg and Spanier \cite{GS} (see also \cite{GS-book}, Ch. 5, Sec. 5.7, and
\cite{CC1}), a subset of $\Z^d$ (resp., $\N^d$) is semi-linear if and only if it is first-order definable in $\cal Z$ (resp., $\cal N$).

  \subsection{Polyhedra}\label{due-2}

We recall only the notions on polyhedra which are used in the sequel, and refer to \cite{Ba-02,Bar-08,Ben,Z95} for more details.
For $A\subseteq\R^d$, we denote by ${\rm dim}(A)$ the dimension of the affine hull ${\rm aff}(A)$ of $A$, and by $\mathring{A}$ the
relative interior of $A$, that is, its interior in ${\rm aff}(A)$ (if ${\rm dim}(A)=d$, this is the interior of $A$). A {\em polyhedron}
is a subset of $\R^d$ of the form
\begin{equation}\label{lm9:eq0-polyhedron}
P = \{{\bf x}\in \R^d \ : \ \ell _i ({\bf x})\geq \gamma_i, \ i\in I\},
\end{equation}
where $I$ is a finite set of indices, each $\ell_i$ is a linear form and $\gamma_i\in\R$; $P$ is {\em rational} if all the
$\ell_i$ and $\gamma_i$ have rational coefficients. A set defined in the same way by finitely many linear inequalities, some of which are
strict, is called a {\em semi-open} polyhedron. If all the $\gamma_i$ are zero, $P$ is a {\em (polyhedral) cone}; a cone is
{\em pointed} if it contains no lines, and a {\em sub-cone} of a cone $K$ is a cone contained in $K$. The {\em conic hull} ${\rm co}(A)$ of
a set $A$ is the set of the linear combinations of elements of $A$ with non-negative coefficients. Every pointed rational cone
$K\neq\{{\bf 0}\}$ is of the form ${\rm co}(A)$ for a finite set $A\subseteq\Z^d$, whose elements are called {\em generators} of $K$, and
$K$ is {\em simplicial} if $A$ can be chosen linearly independent. A {\em face} of $P$ is a set $\{{\bf v}\in P : \pi({\bf v})=\gamma\}$,
where $\pi$ is a linear form such that $\pi\geq\gamma$ on $P$; a {\em vertex} of $P$ is a point of $P$ which is not the midpoint of two
distinct points of $P$; a {\em polytope} is a bounded polyhedron; ${\rm conv}(A)$ is the convex hull of $A$. Finally, the
{\em recession cone} of a non-empty polyhedron $P$ without lines, described as in \eqref{lm9:eq0-polyhedron}, is the pointed cone
\begin{equation}\label{lm9:eq0-reccon}
K_P \ = \ \{ {\bf v}\in \R^d : {\bf p} + \tau {\bf v} \in P \ \mbox{for all} \ {\bf p}\in P \ \mbox{and} \ \tau \geq 0 \} \ = \
\{{\bf x}\in \R^d \ : \ \ell _i ({\bf x})\geq 0, \ i\in I\}.
\end{equation}
The following result describes the structure of $P$ (\cite[Theorem 4.12]{Bar-08}, \cite{VzG-MS-1978}).
\begin{theorem}\label{Bar-Th412}
Let $P$ be a non-empty polyhedron without lines. Then $P$ is the Minkowski sum
$$P \ = \ M + K \ = \ \{{\bf m}+{\bf k} \ : \ {\bf m}\in M, \ {\bf k}\in K\},$$
where $M$ is the convex hull of the (finitely many) vertices of $P$ and $K$ is the recession cone of $P$.
\end{theorem}
All the polyhedra considered in the sequel are rational, even when this is not explicitly written. The {\em indicator function}
$[A]$ of a set $A\subseteq\R^d$ is its characteristic function, and the {\em algebra of rational polyhedra} ${\cal P}(\Q^d)$ is the
$\R$-span of the indicator functions of the rational polyhedra of $\R^d$; it is an algebra under pointwise multiplication, since
$[P_1]\cdot[P_2]=[P_1\cap P_2]$.

We end this section with an elementary lemma.
 \begin{lemma}\label{lm:yuri-neretin-lemma}
Let $P$ be a subset of $\R^d$ defined by finitely many linear inequalities with rational coefficients, each of which
may be strict or not. Then there exists a (closed) rational  polyhedron $\overline P$ such that
$\overline{P} \subseteq P$ and $\overline{P} \cap \Z^d = P \cap \Z^d$. More precisely, $\overline P$ is obtained from $P$ by
replacing each strict inequality $\ell ({\bf x}) > \gamma$ occurring in the description of $P$ with an inequality
$\ell ({\bf x}) \geq \gamma + \varepsilon$, for a suitable $\varepsilon\in\Q$, $\varepsilon >0$; in particular,
$\overline P$ can be computed from $P$.

Moreover, if $P = \q _0 + \mathring{K}$, where $\q _0\in\Q^d$ and $\mathring{K}$ is the relative interior of a simplicial
cone $K$ generated by linearly independent vectors of $\Z^d$, then $\overline{P}$ can be chosen of the form ${\bf q} + K$, with ${\bf q}\in P\cap\Q^d$.
 \end{lemma}

\begin{proof}
Let $\ell({\bf x})>\gamma$ be one of the strict inequalities. Multiplying it by a positive integer, we may assume that $\ell$ has integral
coefficients; then $\ell(\Z^d)\subseteq\Z$, so that, for ${\bf x}\in\Z^d$, one has $\ell({\bf x})>\gamma$ if and only if
$\ell({\bf x})\geq\lfloor\gamma\rfloor+1=\gamma+\varepsilon$, where $\varepsilon=\lfloor\gamma\rfloor+1-\gamma\in(0,1]\cap\Q$. Replacing
the strict inequalities one at a time in this way, one obtains a closed rational polyhedron $\overline P$, which is contained in $P$
since $\varepsilon>0$, and has the same integer points as $P$.

For the second claim, let $K={\rm co}({\bf v}_1,\ldots,{\bf v}_r)$ and let $\ell_1,\ldots,\ell_r,m_1,\ldots,m_{d-r}$ be the basis of the
dual space of $\R^d$ which is dual to a basis ${\bf v}_1,\ldots,{\bf v}_r,{\bf u}_1,\ldots,{\bf u}_{d-r}$ of $\Q^d$. These forms have
rational coefficients, $K=\{{\bf x} : \ell_i({\bf x})\geq 0, \ m_j({\bf x})=0\}$ and
$P=\{{\bf x} : \ell_i({\bf x})>\ell_i(\q_0), \ m_j({\bf x})=m_j(\q_0)\}$. By the first part, one can take
$\overline P=\{{\bf x} : \ell_i({\bf x})\geq\ell_i(\q_0)+\varepsilon_i, \ m_j({\bf x})=m_j(\q_0)\}$, with $\varepsilon_i\in\Q$,
$\varepsilon_i>0$. If $\q$ is the unique solution of the linear system $\ell_i({\bf x})=\ell_i(\q_0)+\varepsilon_i$,
$m_j({\bf x})=m_j(\q_0)$, then $\q$ is rational, $\q\in P$, and $\overline P=\q+K$.
\qed\end{proof}

  \section{Valuations of rational polyhedra}\label{cinque}
  We now describe a connection between two constructions -- both involving valuations of rational polyhedra --  provided, respectively,
   by Brlek and Reutenauer in  \cite{BerReut} and
   by Khovanskii, Pukhlikov and Lawrence in  \cite{Ko-Pu-1992,Law-1990}.
     Let $\bf F$ be a field and let $x_1, \ldots, x_d$ be commutative variables.
Denote by ${\bf F} (x_1, \ldots, x_d)$ the quotient field of the polynomial
    ring ${\bf F} [x_1, \ldots, x_d]$.
  To every rational expression $E$ describing a rational subset of  $\Z^d$,
  one can associate the element $f(x_1, \ldots, x_d)   \in {\bf F} (x_1, \ldots, x_d)$ obtained by replacing
 the symbols of the rational operations in $E$  as follows:
  \begin{itemize}
\item $\cup \longrightarrow +$
\item $S^*\longrightarrow (1 - S)^{-1}$,
 \end{itemize}
while the other symbols are left unchanged.
 In the sequel the mapping
    \begin{equation}\label{eq:def-of-varphi-0}
   E \longrightarrow f(x_1, \ldots, x_d)
     \end{equation}
will be called the {\em BR-coding of $E$.\footnote{The term coding is used here, not to mean an injective monoid morphism but to emphasize the fact that
the mapping is a purely combinatorial transformation of symbols.}}
For instance, if     $E = x^*(y^{-1})^* \cup x^{-1}(x^{-1})^*$ one gets $$f(x,y) = (1-x)^{-1}(1-y^{-1})^{-1} + x^{-1}(1-x^{-1})^{-1}.$$

 The main result of \cite{BerReut} is the following.

             \begin{theorem}\label{th:BR-1}{(\cite{BerReut}, Theorem 1)}
 If $E$ is an unambiguous rational expression, then $f(x_1, \ldots, x_d)$ is a well-defined rational function in
 ${\bf F} (x_1, \ldots, x_d)$, that depends only on the set described by $E$.
  \end{theorem}
  In the sequel we will apply the theorem above in the case ${\bf F}=\R$.
 Recall that  ${\cal P}(\Q^d)$  is   linearly spanned (over $\R$) by the indicator functions of rational polyhedra. 
As a consequence of Theorem \ref{th:BR-1}, one can construct
 a valuation $\varphi$ from ${\cal P}(\Q^d)$ into $\R (x_1, \ldots, x_d)$ having remarkable properties.
 To show this construction, let us state the  Lemma \ref{lm-prs1} below. Such a lemma follows from
 Theorem \ref{ratmGS} and from
 the  characterization   of semi-linear sets in terms of first order definability in Presburger Arithmetic (see Sec. \ref{due-3}).
 \begin{lemma}
\label{lm-prs1} Let $P$ be a rational polyhedron.
  Then the  set  of points of $\Z^{d}$  contained in $P$ is a rational unambiguous subset
  of $\Z^{d}$.
 \end{lemma}
 By Lemma \ref {lm-prs1}, we can define a new map   $\varphi: {\cal P}(\Q^d)\to \R (x_1, \ldots, x_d)$ by extending linearly the map   \begin{equation}\label{eq:def-of-varphi}
   [P] \longrightarrow g(x_1, \ldots, x_d),
     \end{equation}
  where $g(x_1, \ldots, x_d)$ is obtained by applying the BR-coding to an arbitrary unambiguous rational expression $E$ describing $P\cap\Z^d$.
  Note that, by Theorem \ref{th:BR-1}, the map (\ref{eq:def-of-varphi}) is well defined.
    \medskip

    For instance,
      let $P=\{(x_1,x_2)\in\R^2: x_1,x_2\geq 0\}$. Then $P\cap\Z^2=\N^2$ so that
      the rational unambiguous set $P\cap\Z^2 = {\bf e}_1 ^\oplus + {\bf e}_2 ^\oplus,$ with ${\bf e}_1 =(1,0), {\bf e}_2 =(0,1),$ is described by the expression
      $E= ({\bf e}_1)^* ({\bf e}_2)^*$. Then $\varphi([P])=  \frac{1}{(1-x_1)(1-x_2)}$, where such function is        obtained by
      applying the BR-coding to $E$.

 The following theorem holds:

          \begin{theorem}\label{th:BR}{(\cite{BerReut}, Theorem 10)} Let $P$ be a rational polyhedron.
                     \begin{enumerate}

\item[(i)] $\varphi$ is a valuation from ${\cal P}(\Q^d)$ into $\R (x_1, \ldots, x_d)$, that is, the {map}
(\ref{eq:def-of-varphi}) extends to a well-defined linear map on ${\cal P}(\Q^d)$.

\item[(ii)] If $P$ contains some line, then $\varphi([P]) =0$.

\item[(iii)] If $P'=P+{\bf v},\,{\bf v}\in \Z^d$, then
$\;\; \varphi  ([P'])=   {\bf x}^{\bf v}    \varphi ([P]).$
   \end{enumerate}
                \end{theorem}

                                    We now compare Theorem \ref{th:BR}  with the theorem by Khovanskii and Pukhlikov \cite{Ko-Pu-1992} and by Lawrence \cite{Law-1990} stated in the form
reported  in \cite{Bar-08}, Theorem 13.8 (see also  \cite{Ba-02}, Ch. VIII, Theorem 3.3).
To this purpose, recall that, if $P$ is an arbitrary rational polyhedron, the {\em generating series of $P$} is the formal series
\begin{equation}\label{sf}\sum _ {{\bf m}\in P \cap \Z ^d} \ {\bf x}^{\bf m}.\end{equation}
 It is shown in \cite[VIII, Lemma 3.1]{Ba-02} that there exists a non-empty open set $U\subset \mathbb C^d$ such that for all ${\bf x}\in U$ the series
\eqref{sf} converges to a rational function $f(P, {\bf x})$.
\par
Let ${\cal R}(\C^d)$ be the vector space of rational functions on $\C^d$ spanned by the functions of the form $$ \frac{{\bf x}^{\bf v}}{(1- {\bf x}^{{\bf v}_1}) \cdots (1- {\bf x}^{{\bf v}_d}) },$$
where $\bf v$ $\in \Z^d$ and ${\bf v}_i \in \Z^d \setminus \{{\bf 0}\}, \ 1\leq i \leq d$.  The following result holds.

              \begin{theorem}\label{th:KPL}
            There exists a valuation $\mathcal F$ from ${\cal P}(\Q^d)$ to ${\cal R}(\C^d)$ such that the following holds:
             \begin{enumerate}

\item[(i)]

If $P\subset \mathbb R^d$ is a rational polyhedron without straight lines, then $\mathcal F([P]) =
f(P,{\bf x})$ is the rational function such that
$$f(P,{\bf x})=\sum _ {{\bf m}\in P \cap \Z ^d} \ {\bf x}^{\bf m}
$$
provided the series converges absolutely.

\item[(ii)] If $P$ contains some line, then $\mathcal F([P]) \equiv 0$.

\item[(iii)] If $P'$ is the translation of $P$ by $\bf v$ $\in \Z^d$, then
$\mathcal F([P'])=   {\bf x}^{\bf v}\mathcal F([P]).$
\end{enumerate}
                \end{theorem}
\begin{remark}\label{rmk:conv} Let  $K$ be the
recession cone of a (non-empty) rational polyhedron $P$ without lines. Assume that $K$ = co$({\bf w}_1, \ldots, {\bf w}_s)$, for some ${\bf w}_i \in \Z^d \setminus \{{\bf 0}\}, i =1, \ldots, s$.
Then the series \eqref{sf} converges absolutely to $f(P,{\bf x})$ at the points with non-zero coordinates of the non-empty open set
$$W_K = \{ {\bf x}\in \C^d \ : \ |{\bf x} ^ {{\bf w}_i}|< 1, \ i =1, \ldots, s   \}.$$
This follows from Theorem \ref{Bar-Th412}, which gives $P\cap\Z^d\subseteq F+{\bf w}_1^\oplus+\cdots+{\bf w}_s^\oplus$ for a
finite set $F\subseteq\Z^d$: the series is dominated by a finite sum of products of convergent geometric series.
\end{remark}

 The following conjecture relates Theorems       \ref{th:BR} and \ref{th:KPL}.
\medskip

\noindent
{\bf Conjecture  (Brlek and Reutenauer, \cite{BerReut})} {\em The valuations $\varphi$ and $\mathcal F$ coincide.}

     \section{Proof of the Main Theorem}\label{MT}

The proof rests on the following lemma, which reduces the conjecture, for a given polyhedron $P$, to the existence of a suitable
decomposition of the set $P\cap\Z^d$. It will be applied to cones in Section \ref{sec:cones} and to arbitrary polyhedra without lines
in Section \ref{sei:3}.

\begin{lemma}\label{lm:simple-dec}
Let $P$ be a non-empty rational polyhedron without lines and let $K$ be its recession cone. Assume that
\begin{equation}\label{th:FLA-Ankara-crucial-eq0}
P \cap \Z^d \ = \ \bigsqcup _ {i=1} ^ t \ \left({\bf p}_i + {\bf v}_{i,1}^\oplus + \cdots + {\bf v}_{i,r_i}^\oplus\right),
\end{equation}
where, for each $i=1, \ldots, t$, ${\bf p}_i\in\Z^d$ and ${\bf v}_{i,1}, \ldots, {\bf v}_{i,r_i}$ are linearly independent vectors of
$K\cap\Z^d$ (the case $r_i=0$ being allowed, in which case the corresponding piece is the singleton $\{{\bf p}_i\}$), $t\geq 0$, and the union is
disjoint. Then
\begin{equation}\label{th:FLA-Ankara-crucial-eq1}
\varphi ([P]) \ = \ \mathcal F([P]) \ = \
\sum _ {i=1} ^ t \  \frac{{\bf x}^{{\bf p}_i}}{\left(1-{\bf x}^{{\bf v}_{i,1}}\right)\cdots \left(1-{\bf x}^{{\bf v}_{i,r_i}}\right)}.
\end{equation}
\end{lemma}

\begin{proof}
If $t=0$, that is, if $P\cap\Z^d=\emptyset$, then both $\varphi ([P])$ and $\mathcal F([P])$ are zero and there is nothing to prove.
Let $t\geq 1$. By \eqref{th:FLA-Ankara-crucial-eq0}, the set $P\cap\Z^d$ is described by the rational expression
$$E_P \  =  \  \bigcup _ {i=1} ^ t \  {\bf x}^{{\bf p}_i}\left({\bf x}^{{\bf v}_{i,1}}\right)^* \cdots \left({\bf x}^{{\bf v}_{i,r_i}}\right)^*,$$
where, for $r_i =0$, the corresponding term reduces to the monomial ${\bf x}^{{\bf p}_i}$. This expression is unambiguous: the union is
disjoint and, the vectors ${\bf v}_{i,1}, \ldots, {\bf v}_{i,r_i}$ being linearly independent, every element of the $i$-th set is
written in a unique way as ${\bf p}_i + m_1{\bf v}_{i,1} + \cdots + m_{r_i}{\bf v}_{i,r_i}$, with $m_1, \ldots, m_{r_i}\in\N$. Hence, by applying
the BR-coding to $E_P$, one obtains that $\varphi ([P])$ is equal to the right-hand side of \eqref{th:FLA-Ankara-crucial-eq1}.

On the other hand, by \eqref{th:FLA-Ankara-crucial-eq0}, the generating series of $P$,
$f(P, {\bf x})=\sum _ {{\bf m}\in P \cap \Z ^d} \ {\bf x}^{\bf m}$ (seen as a formal series), can be written as
 \begin{equation}\label{th:FLA-Ankara-crucial-eq1-bis}
 \sum _{i=1} ^ t \ \  \left(  \sum _ {{\bf m} \ \in\  {{\bf p}_i + {\bf v}_{i,1}^\oplus + \cdots + {\bf v}_{i,r_i}^\oplus}} \ {\bf x}^{\bf m}\right).
  \end{equation}
Let $K = {\rm co}({\bf w}_1, \ldots, {\bf w}_n)$, ${\bf w}_j \in \Z^d\setminus\{{\bf 0}\}$ (with $n=0$ if $K=\{{\bf 0}\}$), and consider the set
$$W_K^{\times} \ = \ \{ {\bf x}\in \C^d \ : \ x_1\cdots x_d\neq 0, \ \ |{\bf x} ^ {{\bf w}_j}|< 1, \ j =1, \ldots, n \},$$
that is, the set of the points with non-zero coordinates of the set $W_K$ of Remark \ref{rmk:conv}; since $W_K$ is open and not empty, so
is $W_K^{\times}$. Every vector ${\bf v}_{i,k}$ occurring in \eqref{th:FLA-Ankara-crucial-eq0} is a non-zero vector of $K$, say
${\bf v}_{i,k} = \sum _{j} \lambda _j {\bf w}_j$ with $\lambda _j \geq 0$ not all zero; hence, for every ${\bf x}\in W_K^{\times}$,
$$\left|{\bf x} ^ {{\bf v}_{i,k}}\right| \ = \ \prod _j \left|{\bf x} ^ {{\bf w}_j}\right| ^{\lambda _j} \ < \ 1 .$$
Consequently, for each $i$, the series
$\sum _ {{\bf m} \in {\bf p}_i + {\bf v}_{i,1}^\oplus + \cdots + {\bf v}_{i,r_i}^\oplus} {\bf x}^{\bf m}$ is the product of the monomial
${\bf x}^{{\bf p}_i}$ and of $r_i$ absolutely convergent geometric series, and it converges absolutely on $W_K^{\times}$ to
$${\bf x}^{{\bf p}_i}\left(1-{\bf x}^{{\bf v}_{i,1}}\right)^{-1}\cdots \left(1-{\bf x}^{{\bf v}_{i,r_i}}\right)^{-1}.$$
Therefore the series \eqref{th:FLA-Ankara-crucial-eq1-bis}, that is $f(P, {\bf x})$, converges absolutely on $W_K^{\times}$ to the
right-hand side of \eqref{th:FLA-Ankara-crucial-eq1}. By Theorem \ref{th:KPL}--(i), the latter rational function is $\mathcal F([P])$
(two rational functions which coincide on a non-empty open subset of $\C^d$ are equal).
\qed\end{proof}

\subsection{The case of cones}\label{sec:cones}

   In the sequel, cones will be always assumed to be pointed.
    Let $K=${co}$({\bf v}_1, \ldots, {\bf v}_\ell)$ be a simplicial cone generated by linearly independent vectors ${\bf v}_1, \ldots, {\bf v}_\ell$ in
    $\Q ^d$.
        The {\em fundamental parallelepiped of ${K}$} is the (semi-open) polytope
    $\Pi _ {K} = \left\{\sum _ {i=1} ^ \ell \ \alpha _i {\bf v}_i \ : \ 0 \leq \alpha_i < 1, \ i=1, \ldots, \ell \right \}.$

    In the sequel we simply denote the fundamental parallelepiped by $\Pi = \Pi _ { K}$, and associate with it the Laurent polynomial:
 \begin{equation}\label{eq:pol-fund-par}
  p_{\Pi} ({\bf x})=\sum _{{\bf u} \in\  \Pi   \ \cap \  \Z^d}\  \ {\bf x}^{\bf u}.
\end{equation}

The following proposition is well known (see, e.g., \cite{Stanley82} and \cite{Bar-08}).
 \begin{proposition}\label{cor: transl-of-a-cone-simpl}
                    Let ${ K} = $ {co}(${\bf v}_1, \ldots, {\bf v}_\ell)$ be a simplicial cone
generated by
linearly independent vectors  ${\bf v}_1, \ldots, {\bf v}_\ell\in\Z^d$, and let $\Pi$ be its
 fundamental parallelepiped.    Let $\q \in \Q^d$ be an arbitrary vector.  Then one has
 \begin{equation}\label{eq:cor: transl-of-a-cone-simpl}
     (\q+{K}) \cap \Z^d \  = \  \bigsqcup _ {{\bf p}\, \in \, (\q \, +\,  \Pi) \, \cap\,  \Z^d} \  \ {\bf p}  + {\bf v_1}^\oplus + \cdots  +{\bf v_\ell}^\oplus,
\end{equation}
and the union is disjoint.
\end{proposition}
Indeed, every ${\bf u}\in(\q+K)\cap\Z^d$ is written in a unique way as ${\bf u}=\q+\sum_{i}\alpha_i{\bf v}_i$ with
$\alpha_i\geq 0$, and then ${\bf u}={\bf p}+\sum_i\lfloor\alpha_i\rfloor{\bf v}_i$, where
${\bf p}=\q+\sum_i(\alpha_i-\lfloor\alpha_i\rfloor){\bf v}_i$ is the unique point of $(\q+\Pi)\cap\Z^d$ such that
${\bf u}\in{\bf p}+{\bf v}_1^\oplus+\cdots+{\bf v}_\ell^\oplus$; the set $(\q+\Pi)\cap\Z^d$ is finite, since $\Pi$ is bounded.

The next proposition describes the integer points of an arbitrary pointed rational cone with rational apex as a disjoint union of
simple sets whose apexes are \emph{integral}; it is the form in which the decomposition of cones will be used in the sequel.

 \begin{proposition}\label{prop-refinement-cones}
Let $C$ be a pointed rational cone and let ${\bf q}\in\Q^d$. Then
\begin{equation}\label{eq:prop-refinement-cones}
(\q + C)\cap\Z^d \ = \ \bigsqcup _ {i=1} ^ {a} \ \left({\bf p}_i + {\bf v}_{i,1}^\oplus + \cdots + {\bf v}_{i,r_i}^\oplus\right),
\end{equation}
where, for each $i=1, \ldots, a$, ${\bf p}_i\in\Z^d$ and ${\bf v}_{i,1}, \ldots, {\bf v}_{i,r_i}$ are linearly independent vectors of
$C\cap\Z^d$ (the case $r_i=0$ being allowed, in which case the corresponding piece is $\{{\bf p}_i\}$), and the union is disjoint. In
particular, $(\q + C)\cap\Z^d$ is a rational unambiguous set. Moreover, the decomposition \eqref{eq:prop-refinement-cones} is computable
from $\q$ and from a finite set of generators of $C$.
\end{proposition}

\begin{proof}
If $C=\{{\bf 0}\}$, then $(\q+C)\cap\Z^d$ is either empty or equal to $\{\q\}$. Otherwise, $C={\rm co}({\bf w}_1,\ldots,{\bf w}_n)$ with
${\bf w}_1,\ldots,{\bf w}_n\in\Z^d$, and $C$ admits a triangulation without new rays: $C=C_1\cup\cdots\cup C_s$, where each $C_i$ is a
simplicial cone generated by linearly independent vectors among ${\bf w}_1,\ldots,{\bf w}_n$, and any two of these cones meet in a common
face (\cite{Ba-02}, Ch. VIII, Lemmas 2.2 and 2.3). Recall that a simplicial cone ${\rm co}({\bf v}_1,\ldots,{\bf v}_r)$ is the disjoint
union of $\{{\bf 0}\}$ and of the relative interiors of its faces of positive dimension: the face containing a point
${\bf x}=\sum_i\lambda_i{\bf v}_i\neq{\bf 0}$ in its relative interior is the cone generated by the vectors ${\bf v}_i$ with
$\lambda_i>0$ (cf \cite{Bar-08}, Theorem 4.15). Let $\{\tau_j\}_{j\in J}$ be the family of the faces of positive dimension of the cones
$C_1,\ldots,C_s$, each face being listed once. Their relative interiors $\mathring{\tau}_j$ cover $C\setminus\{{\bf 0}\}$, and they are
pairwise disjoint: if ${\bf x}\in\mathring{\tau}\cap\mathring{\tau}'$, where $\tau$ and $\tau'$ are faces of $C_i$ and of $C_k$, then
$\tau$ and $\tau'$, being the smallest faces of $C_i$ and $C_k$ containing ${\bf x}$, are faces of the simplicial cone $C_i\cap C_k$, so
that $\tau=\tau'$. Hence
$$(\q + C)\cap\Z^d \ = \ \left(\{\q\}\cap\Z^d\right) \ \sqcup \ \bigsqcup _ {j\in J} \ \left((\q + \mathring{\tau}_j)\cap\Z^d\right).$$
By the second part of Lemma \ref{lm:yuri-neretin-lemma}, $(\q + \mathring{\tau}_j)\cap\Z^d=(\q'_j + \tau_j)\cap\Z^d$ for a computable
point $\q'_j\in\Q^d$, and Proposition \ref{cor: transl-of-a-cone-simpl} decomposes the latter set into finitely many pairwise disjoint
simple sets with integral apexes, whose periods are the generators of $\tau_j$, which lie in $C\cap\Z^d$. All the steps are effective.
\qed\end{proof}

We can now solve the conjecture for cones and, more generally, for their translates by rational vectors.

\begin{proposition}\label{prop:BR-affine-cones}
Let $C$ be a pointed rational cone and let ${\bf q}\in\Q^d$. Then $\varphi ([\q + C]) = \mathcal F([\q + C])$. In particular, the
conjecture of Brlek and Reutenauer holds for cones.
\end{proposition}

\begin{proof}
The set $\q + C$ is a rational polyhedron without lines, whose recession cone is $C$. By Proposition \ref{prop-refinement-cones},
the set $(\q + C)\cap\Z^d$ admits the decomposition \eqref{eq:prop-refinement-cones} into pairwise disjoint simple sets, with integral
apexes and with all the periods in $C\cap\Z^d$. Hence the claim follows from Lemma \ref{lm:simple-dec}.
\qed\end{proof}

\begin{remark}\label{rmk:simplicial}
If $K = {\rm co}({\bf v}_1, \ldots, {\bf v}_\ell)$ is a simplicial cone generated by linearly independent vectors of $\Z^d$ and
$\q\in\Q^d$, then Proposition \ref{cor: transl-of-a-cone-simpl} already provides a decomposition of $(\q + K)\cap\Z^d$ of the form
required by Lemma \ref{lm:simple-dec}, which therefore gives the explicit formula
$$\varphi ([\q + K]) \ = \ \mathcal F([\q + K]) \ = \
\frac{\sum _{{\bf p}\, \in \, (\q \, +\,  \Pi) \, \cap\,  \Z^d} \ {\bf x}^{\bf p}}{(1- {\bf x}^{{\bf v}_1}) \cdots (1- {\bf x}^{{\bf v}_\ell})}.$$
For $\q = {\bf 0}$ the numerator is the polynomial $p_{\Pi} ({\bf x})$ of Eq. (\ref{eq:pol-fund-par}) (cf also \cite{Bar-08}, Lemma 13.5).
\end{remark}

      \subsection{The general case}\label{sei:3}

To deal with the general case, we need the following theorem, which provides an
  unambiguous version of Theorem~\ref{Bar-Th412}.
Its proof, which is based on the key technical  Proposition \ref{prop:MainDecomp},  is deferred to Section \ref{sei:4}.

       \begin{theorem}\label{cor:MainDecomp}
Let $P$ be a rational polyhedron without lines and let $K$ be its recession cone. There exists an algorithm that computes a
rational unambiguous decomposition of the set $P \cap \Z^d$
         \begin{equation}\label{eq-main:cor:MainDecomp}
       P \cap \Z^d \ = \ \bigsqcup _ {i=1} ^ t \ \left({\bf p}_i + {\bf v}_{i,1}^\oplus + \cdots + {\bf v}_{i,r_i}^\oplus\right),
       \end{equation}
where, for each $i=1, \ldots, t$, ${\bf p}_i\in\Z^d$ and ${\bf v}_{i,1}, \ldots, {\bf v}_{i,r_i}$ are linearly independent vectors of
$K\cap\Z^d$ (the case $r_i=0$ being allowed, in which case the corresponding piece is the singleton $\{{\bf p}_i\}$). In other words,
\eqref{eq-main:cor:MainDecomp} is a decomposition of $P\cap\Z^d$ into pairwise disjoint simple sets, with integral apexes and with all
the periods lying in the recession cone of $P$.
 \end{theorem}

      Notice that the proof of Theorem \ref{th:MAIN-B} below uses only the existence of the decomposition
\eqref{eq-main:cor:MainDecomp} (see also Remark \ref{rmk:ES}).
Now we can state

      \begin{theorem}\label{th:MAIN-B}{(Main Theorem)}
    The valuations $\mathcal F$  and $\varphi$ coincide on the algebra ${\cal P}(\Q^d)$.
     \end{theorem}

     \begin{proof}
 Since $\mathcal F$ and $\varphi$ are valuations, it is enough to prove that,
     for every rational polyhedron $P$, one has $\mathcal F([P])=\varphi ([P])$. 
       If $P$ is a polytope, then $P\cap \Z^d$ is finite so that $\mathcal F([P])$ 
       and $\varphi ([P])$ are
     Laurent polynomials that, by construction, coincide.
      Assume that $P$ is not bounded. If $P$ contains some line, then the claim follows from Theorem \ref{th:BR}--(ii) and Theorem  \ref{th:KPL}--(ii).
          If $P$ is not bounded and does not contain lines, then, by Theorem \ref{cor:MainDecomp}, the set $P \cap \Z^d$ admits the
partition (\ref{eq-main:cor:MainDecomp}) into pairwise disjoint simple sets, with integral apexes and with all the periods lying in the
recession cone of $P$. Thus the hypothesis of Lemma \ref{lm:simple-dec} is satisfied, and
$\varphi ([P]) = \mathcal F([P])$ by \eqref{th:FLA-Ankara-crucial-eq1}.

    \qed  \end{proof}

\begin{remark}\label{rmk:ES}
The proof of Theorem \ref{th:MAIN-B} uses Theorem \ref{cor:MainDecomp} only through the existence of a decomposition of the form
\eqref{eq-main:cor:MainDecomp}. Such a decomposition is also provided by the theorem of Eilenberg and Sch\"utzenberger: by
Lemma \ref{lm-prs1} and Theorem \ref{ratmGS}, the set $P\cap\Z^d$ is a finite disjoint union of simple sets
${\bf p}_i + {\bf v}_{i,1}^\oplus + \cdots + {\bf v}_{i,r_i}^\oplus$, and every period ${\bf v}_{i,k}$ lies in the recession cone of
$P$, because ${\bf p}_i+n{\bf v}_{i,k}\in P$ for every $n\in\N$ and $P$ is convex. Unlike Theorem \ref{cor:MainDecomp}, however, this
argument does not exhibit the decomposition, nor the corresponding rational expression, in terms of the inequalities defining $P$.
\end{remark}

       \section{Proof of Theorem \ref{cor:MainDecomp}}\label{sei:4}
       {The proof of Theorem \ref{cor:MainDecomp} requires some intermediate  facts.
            In order to state them, we fix some notation. }
                         Recall that, by Theorem \ref{Bar-Th412} ,
$
P = M + K$,
      where $K$ is the recession cone of $P$ and $M$ is the polytope given by the convex hull of the vertices of $P$. \par
 In the following discussion, we represent the polyhedron $P$ and its recession cone $K$ as
\begin{equation}\label{eq:main-P-and-K}
 P = \{{\bf x}\in \R^d \ : \ell _i ({\bf x})\geq \alpha_i, \ i\in I\},\quad
K = \{{\bf x}  \in \R^d \ :  \ \ell _i ({\bf x})\geq 0, \ i\in I\},
\end{equation}
respectively.
We shall use the following two facts.
\begin{enumerate}
\item[\bf Fact 1.] 
 Let ${\bf w}\in \R^d$ be an arbitrary vector and let
   $\tau _ {\bf w} : \R^d \to\R^d$ denote the translation by ${\bf w}$.

                                 If {dim}$(K) = d$,
             there exists  ${\bf w}\in \Z^d$ such that $\tau _ {\bf w} (P)$ is contained in the interior $\mathring{K}$ of $K$.
Indeed, the open cone $\mathring{K}$ contains a vector ${\bf u}\in\Z^d$; since $M$ is bounded and
$\ell_i({\bf u})>0$ whenever $\ell_i\neq 0$, one has $M+n{\bf u}\subseteq\mathring{K}$ for some positive integer $n$, and then
$\tau _{n{\bf u}}(P)=(M+n{\bf u})+K\subseteq\mathring{K}+K=\mathring{K}$.

 \item[\bf Fact 2.] \label{prop:MainDecomp-lm-cone}

Assume that ${\rm dim}(K)=d$, fix a point $\q$ in the (non-empty) interior $\mathring{K}$ of $K$, and set
$\gamma_i=\ell_i(\q)$, $i\in I$, so that
\begin{equation}\label{prop:MainDecomp-rmk-cone:eq1}
\q + K = \{{\bf x}  \in \R^d \ :  \ \ell _i ({\bf x})\geq \gamma_i, \ i\in I\}.
\end{equation}
With every ${\bf x}\in K$ associate the word $(f_i({\bf x}))_{i\in I}$ over the alphabet $\{+,-\}$ (with respect to a fixed enumeration
of $I$), where $f_i({\bf x})=+$ if $\ell_i({\bf x})\geq\gamma_i$ and $f_i({\bf x})=-$ otherwise, and write ${\bf x}\sim{\bf x}'$ if
${\bf x}$ and ${\bf x}'$ have the same word. Every $\sim$-class is a semi-open polyhedron, and the class of the word $(+,\ldots,+)$ is
$\q+K$. Hence (see {\bf Fig.} 1) the cone $K$ admits the partition into finitely many sets
\begin{equation}\label{prop:MainDecomp-lm-cone-eq}
K  \  = \  (\q + K)\  \cup \  \bigcup _ {C\in \Omega}\  C,
\end{equation}
where $\Omega$ is the set of the $\sim$-classes whose word contains at least one occurrence of the symbol $'-'$.

\begin{center}
\setlength{\unitlength}{.2mm}
\begin{picture}(290,345)(70,290)
  \put(80, 300){{\bf Fig.  $1$}}
 \put(198,471){{\bf q} }   \put(95,373){$\bf 0$ }    \tiny{ \put(285, 575){$ (+,+)$}  \put(200, 575){$(-,+)$}  \put(285, 490){$(+,-)$}  \put(130, 430){$(-,-)$}       }
   \put(195,483){\vector(1,2){65}}
    \put(195,483){\vector(3,1){155}}
   \put(100,391){\vector(1,2){120}}
   \put(100,391){\vector(3,1){255}}
    \put(100,391){\circle*{3}}
     \put(195,483){\circle*{3}}
\put(195,483){\line(-1,-2){36}} \put(195,483){\line(-3,-1){57}}
\end{picture}
\end{center}

 \end{enumerate}
The following lemma describes the recession cones of the closed polyhedra which, in the proof of
Proposition \ref{prop:MainDecomp}, replace the pieces of the partition of Fact $2$.
\begin{lemma}\label{constr-rec-con-Rv}
Let $P$ and $K$ be as in \eqref{eq:main-P-and-K}. Assume that ${\rm dim}(K)=d$ and $P\subseteq \mathring{K}$, let $\q\in P\cap\Q^d$,
and consider the partition \eqref{prop:MainDecomp-lm-cone-eq} of $K$ with respect to $\q$. Let $C\in\Omega$, and denote by $I_-$
(resp., $I_+$) the set of the indices $i\in I$ such that the word defining $C$ has the symbol $'-'$ (resp., $'+'$) at position $i$,
so that $I_-\neq\emptyset$. Then
\begin{equation}\label{eq:PcapC}
P\cap C \ = \ \{{\bf x}\in P \ : \ \ell _i ({\bf x})\geq \gamma_i, \ i \in I_+, \ \ \ \ell _i ({\bf x}) < \gamma_i, \ i \in I_- \}.
\end{equation}
Moreover, let $\varepsilon _i >0$, $i\in I_-$, and assume that the closed polyhedron
\begin{equation}\label{eq:RC}
R_C \ = \ \{{\bf x}\in P \ : \ \ell _i ({\bf x})\geq \gamma_i, \ i \in I_+, \ \ \ \ell _i ({\bf x}) \leq \gamma_i - \varepsilon _i, \ i \in I_- \}
\end{equation}
is not empty. Then the recession cone of $R_C$ is
$$K_C \ = \ \{{\bf x}\in K \ : \ \ell _i ({\bf x}) = 0, \ i \in I_- \},$$
that is, the face of $K$ on which the linear forms $\ell _i$, $i\in I_-$, vanish. In particular, $0\leq {\rm dim}(K_C)\leq d-1$.
\end{lemma}

\begin{proof}
By the definition of the equivalence $(\sim)$, one has
$C = \{{\bf x}\in K : \ell _i ({\bf x})\geq \gamma_i, \ i \in I_+, \ \ell _i ({\bf x}) < \gamma_i, \ i \in I_- \}$, and
\eqref{eq:PcapC} follows, since $P\subseteq K$. Recall that $\gamma _i = \ell _i(\q)\geq \alpha _i$ for every $i\in I$, because $\q\in P$.
Hence, for $i\in I_+$, the inequality $\ell _i ({\bf x})\geq \alpha _i$ occurring in the description of $P$ is implied by
$\ell _i ({\bf x})\geq \gamma_i$, so that
$$R_C \ = \ \{{\bf x}\in \R^d \ : \ \ell _i ({\bf x})\geq \gamma_i, \ i \in I_+, \ \ \ \alpha _i \leq \ell _i ({\bf x}) \leq \gamma_i - \varepsilon _i, \ i \in I_- \}.$$
By \eqref{lm9:eq0-reccon}, the recession cone of $R_C$ is therefore
$$\{{\bf x}\in \R^d \ : \ \ell _i ({\bf x})\geq 0, \ i \in I_+, \ \ \ \ell _i ({\bf x}) = 0, \ i \in I_- \} \ = \ K_C .$$
Each linear form $\ell _i$, $i\in I$, is non-negative on $K$, so that $\{{\bf x}\in K : \ell _i ({\bf x})=0\}$ is a face of $K$;
being an intersection of faces, $K_C$ is a face of $K$. Finally, let $j\in I_-$. The form $\ell _j$ is not zero, because
$\ell _j({\bf x}) < \ell _j (\q)$ for every ${\bf x}\in R_C$, and $K_C$ is contained in the hyperplane $\{\ell _j = 0\}$, whence
${\rm dim}(K_C)\leq d-1$.
\qed\end{proof}

We can now state the proposition which provides our main technical tool.

  \begin{proposition}\label{prop:MainDecomp}
                           Let $P$ be a rational unbounded polyhedron in $\R^d$ without lines, given by
                           the Minkowski sum $P = M + K$, where $M$ is the convex envelope of the vertices of $P$ and  $K$ is the recession cone of $P$.
                                             Then there exist a family  $\{ {\bf q} + K _  {\q}\}_{{\bf q} \in \Gamma},$
where $\Gamma \subseteq P \cap \Q^d$ is a  set of rational points of $P$, and each $K _{\q}$ is a
subcone of $K$, such that the following properties hold:
                          \begin{itemize}
 \item[(1)]
            \begin{equation}\label{prop:MainDecomp-eq0}
P \cap \Z^d = \   \bigsqcup _ {{\bf q} \in \Gamma} \left(({\bf q} + K _  {\q})\cap\Z^d\right),
\end{equation}
\noindent
 where the family  $\{ {\bf q} + K _  {\q}\}_{{\bf q} \in \Gamma}$ consists of pairwise disjoint sets.

 \item[(2)]  $\Gamma$ is finite, and each $K_\q$ is a face of $K$.
   \end{itemize}
Moreover, the family $\{ {\bf q} + K _  {\q}\}_{{\bf q} \in \Gamma}$ can be computed from the inequalities defining $P$ (see
Remark \ref{rmk-Prop6}).
   \end{proposition}

      \begin{proof}
We prove (1) and (2) simultaneously, by induction on ${\rm dim}(P)$. If ${\rm aff}(P)\cap\Z^d=\emptyset$, one takes
$\Gamma=\emptyset$. Otherwise, since ${\rm aff}(P)$ is defined by linear equations with rational coefficients, there is an affine
isomorphism of ${\rm aff}(P)$ onto $\R^k$, $k={\rm dim}(P)$, which maps ${\rm aff}(P)\cap\Z^d$ onto $\Z^k$ (it is given by an
integer point of ${\rm aff}(P)$ and a basis of the corresponding lattice); it preserves rational points, recession cones and their
faces. Hence we may assume that ${\rm dim}(P)=d$, and the same reduction is tacitly applied to the polyhedra of smaller dimension which
occur below. If $d=0$, the claim is trivial.
If $d=1$, then $P= M+K,$ where
  $K=$co$({\bf v})$ with ${\bf v} \in \Z^d\setminus \{{\bf 0}\}$  and we can assume that $M= {\{\bf m \}},\ {\bf m}\in \Q ^ d$.  Then $P$ is the translate of a cone and the claim is satisfied by taking
 $ \Gamma = \{ \q\},$ with $\q = {\bf m}, \; K_{\q}=K.$

  We now deal with the induction step. There are two cases, according to the dimension of $K$.

  \medskip

  \noindent

     {\bf Case 1.} dim$(K) \leq d-1$. Since ${\rm dim}(K)<d$, there exists a non-zero linear form $\rho$ with rational coefficients
which vanishes on $K$. For $\beta\in\R$ set $H^\rho_\beta=\{{\bf x}\in\R^d : \rho({\bf x})=\beta\}$, so that $K\subseteq H^\rho_0$.
Since $\rho$ vanishes on $K$, one has $\rho(P)=\rho(M+K)=\rho(M)$, which is a closed interval $[\beta_0,\beta_1]$ with
$\beta_0,\beta_1\in\Q$, $M$ being a rational polytope. Hence
  \begin{equation}\label{prop:MainDecomp-eq1-bis}
P\  = \   \bigsqcup _ {\beta_0\, \leq \beta \leq \, \beta_1}\  {P_\beta}, \quad\quad P_\beta= P \cap H^\rho_\beta .
\end{equation}
Since $\rho$ has rational coefficients, $\rho(\Z^d)$ is a non-zero finitely generated subgroup of $(\Q,+)$, hence
$\rho(\Z^d)=c\,\Z$ for some rational $c>0$. Therefore the set $B=\rho(\Z^d)\cap[\beta_0,\beta_1]$ is finite, and
  \begin{equation}\label{prop:MainDecomp-eq1}
P\cap \Z^d\  = \   \bigsqcup _{\beta\in B}\ \left(P_\beta\cap\Z^d\right).
\end{equation}
Let $\beta\in B$ be such that $P_\beta\neq\emptyset$. Then $P_\beta$ is a rational polyhedron without lines, since $P_\beta\subseteq P$;
its recession cone is $K\cap H^\rho_0=K$, so that $P_\beta$ is unbounded; and ${\rm dim}(P_\beta)\leq d-1$. Hence we can apply the
inductive hypothesis to $P_\beta$: there is a finite set $\Gamma_\beta\subseteq P_\beta\cap\Q^d$ such that
  \begin{equation}\label{prop:MainDecomp-eq1-rmkPaolo}
P_\beta \cap \Z^d \ = \ \bigsqcup _ {{\bf q} \in \Gamma_\beta} \left(({\bf q} + K _  {\q})\cap\Z^d\right),
\end{equation}
where the sets ${\bf q} + K_\q$, ${\bf q}\in\Gamma_\beta$, are pairwise disjoint and each $K_\q$ is a face of $K$. By
\eqref{prop:MainDecomp-eq1}, setting $\Gamma=\bigcup_{\beta\in B}\Gamma_\beta$, which is a finite subset of $P\cap\Q^d$, one gets
\begin{equation}\label{prop:MainDecomp-eq1-rmkPaolo-bis}
P\cap \Z^d \ = \ \bigsqcup _ {{\bf q} \in \Gamma} \left(({\bf q} + K _  {\q})\cap\Z^d\right).
\end{equation}
The sets ${\bf q} + K_\q$, ${\bf q}\in\Gamma$, are pairwise disjoint: for ${\bf q}\in\Gamma_\beta$ one has
${\bf q}+K_\q\subseteq H^\rho_\beta$, because $K_\q\subseteq K\subseteq H^\rho_0$, and the hyperplanes $H^\rho_\beta$, $\beta\in B$, are
pairwise disjoint. Thus \eqref{prop:MainDecomp-eq1-rmkPaolo-bis} affords the decomposition (\ref{prop:MainDecomp-eq0}) for $P$
satisfying $(1)$.

    \bigskip

  \noindent
  {\bf Case 2. dim$(K) = d$.}
 \medskip

 \noindent
By Fact $1$, after a translation by a vector of $\Z^d$, which preserves integer points, lines, recession cones and
their faces, we may assume that $P\subseteq\mathring{K}$.
  Fix a point $\bf q$ in $M\cap \Q^d$. 
  Since ${\bf q}\in M\subseteq P\subseteq \mathring{K}$,
 by Fact $2$, 
 $K$ admits the finite partition (\ref{prop:MainDecomp-lm-cone-eq}) with respect to $\bf q$.
Taking into account that ${\bf q} + K \subseteq P$, one thus gets the partition
\begin{equation}\label{prop:MainDecomp-eq3}
P \  = \  (\q + K)\  \cup \  \bigcup _ {C\in \Omega}\  P \cap C.
 \end{equation}
 Take an arbitrary polyhedron $P \cap C$ in the partition (\ref{prop:MainDecomp-eq3}) distinct from $\q+K$.
 Since $P \cap C$ is semi-open,
 by Lemma \ref{lm:yuri-neretin-lemma}, we can replace $P \cap C$
 with a closed sub-polyhedron $R_C$ of $P \cap C$ such that $R_C\cap\Z^d= P \cap C\cap\Z^d$.
   By Lemma \ref{lm:yuri-neretin-lemma} and by \eqref{eq:PcapC}, the polyhedron $R_C$ has the form \eqref{eq:RC}, for suitable
   rational numbers $\varepsilon _i >0$, $i\in I_-$. Hence, by Lemma \ref{constr-rec-con-Rv}, if $R_C$ is not empty, then its recession
   cone is a face of $K$ of dimension at most $d-1$.

  Let now $R$ be the set
 \begin{equation}\label{prop:MainDecomp-eq3-ter}
R \  = \  (\q + K)\  \cup \  \bigcup _ {C \in \Omega}\  R_C,
 \end{equation}
 obtained by replacing in (\ref{prop:MainDecomp-eq3}) each  $P \cap C$ with $R_C$.
Since  (\ref{prop:MainDecomp-eq3}) is a finite partition and, for each $C\in \Omega,$  $R_C \subseteq P \cap C$,
then  (\ref{prop:MainDecomp-eq3-ter}) defines a finite partition as well. Moreover $R \cap \Z^d= P \cap \Z^d$.
Let us consider now a non-empty polyhedron $R_C$ from (\ref{prop:MainDecomp-eq3-ter}) (for the empty ones we set $\Gamma_C=\emptyset$).
    By Theorem \ref{Bar-Th412}, one has
$$R_C = M_C + K_C,$$
  where $M_C$ is the convex hull of  the vertices of $R_C$ and $K_C$ is the recession cone of $R_C$.

  If dim$(K_C)=0$, then the scheme does not further decompose $R_C$ and it keeps $R_C$ as it is. Indeed,
  $R_C$ is a polytope, so that $R_C\cap\Z^d$ is a finite set, and one can write
  $R_C\cap\Z^d = \bigsqcup _ {{\bf q}' \in \Gamma_C} \left(({\bf q}' + K _  {\q '})\cap\Z^d\right)$,
  where $\Gamma_C = R_C\cap \Z^d$ and $K _  {\q '} = \{ {\bf 0} \}$ for each $\q '\in \Gamma_C$.

   Now suppose that  dim$(R _C)\leq d-1$ and dim$(K_C)>0$, and let us verify that $R_C$ fulfils the induction hypothesis.
   Since $R_C \subseteq P$, then $R_C$ is without lines. Moreover $R_C$ is unbounded, since ${\rm dim}(K_C)>0$.
   So we can apply  to $R_C$ the inductive hypothesis. Thus
 there exists a decomposition (\ref{prop:MainDecomp-eq0}) for $R_C$ satisfying $(1)$ and $(2)$, such that
 \begin{equation}\label{prop:MainDecomp-eq3-ter-2}
 R_C \cap \Z^d = \
  \bigsqcup _ {{\bf q}' \in \Gamma_C} \left(({\bf q}' + K _  {\q '})\cap\Z^d\right),
   \end{equation}
where
  \begin{equation}\label{prop:MainDecomp-eq3-ter-2-bis}
  \Gamma_C \subseteq  R_C \cap \Q^d  \subseteq P \cap \Q^d.
 \end{equation}

Finally, assume that dim$(R _C)= d$ and dim$(K_C)>0$. Since dim$(K_C)\leq d-1$ by Lemma \ref{constr-rec-con-Rv}, the polyhedron
$R_C$ falls under Case $1$, which has already been settled for all the polyhedra of dimension $d$ (its proof uses the inductive
hypothesis only for polyhedra of dimension at most $d-1$). Hence there exists the same decomposition (\ref{prop:MainDecomp-eq3-ter-2})   for $R_C$
satisfying (\ref{prop:MainDecomp-eq3-ter-2-bis}).

Plugging (\ref{prop:MainDecomp-eq3-ter-2}) into (\ref{prop:MainDecomp-eq3-ter}) and since $P \cap \Z^d = R \cap \Z^d$, one gets
  \begin{equation}\label{prop:MainDecomp-eq3-ter-3}
 P \cap \Z^d \  = \  (\q + K) \cap \Z^d\ \  \cup\  \  \bigcup _ {C \in \Omega }    \ \bigcup _ {{\bf q}' \in \Gamma_C} \left(({\bf q}' + K _  {\q '})\cap\Z^d\right),
 \end{equation}
 that is,
 \begin{equation}\label{prop:MainDecomp-eq3-ter-4}P \cap \Z^d \  =  \  \bigcup _ {{\bf q}' \in \Gamma} \left(({\bf q}' + K _  {\q '})\cap\Z^d\right),
  \end{equation}
 where $K_\q = K$ and
 $$\Gamma \ = \  \{\q\}\ \cup\  \bigcup _ {C \in \Omega }    \  \Gamma_C \subseteq P \cap \Q ^d$$
(recall that $\Gamma_C\subseteq R_C\subseteq P$ for every $C\in\Omega$). Thus (\ref{prop:MainDecomp-eq3-ter-4}) has the form of
the decomposition (\ref{prop:MainDecomp-eq0}), and it remains to show that its members are pairwise disjoint.
For every $C\in\Omega$ and every $\q '\in\Gamma_C$ one has
$\q ' + K_{\q '}\subseteq R_C$, because $\q '\in R_C$ and $K_{\q '}$ is contained in the recession cone $K_C$ of $R_C$; the sets $R_C$,
$C\in\Omega$, are pairwise disjoint and disjoint from $\q + K = \q + K_\q$, since so are the sets of the partition (\ref{prop:MainDecomp-eq3}); and,
for a fixed $C$, the sets $\q ' + K_{\q '}$, $\q '\in\Gamma_C$, are pairwise disjoint by construction. This proves condition (1) also
in Case $2$.

\medskip

\noindent
It remains to check that condition (2) holds at each step of the induction. In the base case, $\Gamma=\{{\bf m}\}$ and $K_{\bf m}=K$. In Case $1$,
$\Gamma=\bigcup_{\beta\in B}\Gamma_\beta$ is a finite union of finite sets, and each $K_\q$ is a face of $K$ by the inductive
hypothesis. In Case $2$, the set of apexes is $\{\q\}\cup\bigcup_{C\in\Omega}\Gamma_C$, where $\Omega$ is finite (it has at most
$2^{\card(I)}$ elements) and each $\Gamma_C$ is finite: it is empty if $R_C=\emptyset$; if ${\rm dim}(K_C)=0$, then $\Gamma_C=R_C\cap\Z^d$ is the set of integer points of
a polytope; otherwise, $\Gamma_C$ is finite by the inductive hypothesis or, when ${\rm dim}(R_C)=d$, by Case $1$. Moreover, the cone
attached to $\q$ is $K$, while the cones attached to the points of $\Gamma_C$ are $\{{\bf 0}\}$ when ${\rm dim}(K_C)=0$ and, otherwise,
faces of $K_C$, which is in turn a face of $K$ by Lemma \ref{constr-rec-con-Rv}. Since $\{{\bf 0}\}$ is a face of the pointed cone $K$,
and a face of a face of $K$ is a face of $K$, every cone of the family is a face of $K$. Finally, the translation by an integral vector used in
Case $2$ (Fact $1$) does not affect these properties. Condition (2) is proved.
  \qed       \end{proof}

\begin{remark}\label{rmk-Prop6}
The proof of Proposition~\ref{prop:MainDecomp} is effective: all the objects involved can be computed from the inequalities defining
$P$ (vertices, recession cones and affine hulls of rational polyhedra; an integer point of an affine hull and a basis of the corresponding
lattice; a rational point of $M$; the at most $2^{\card(I)}$ chambers of Fact~$2$; the closed polyhedra given by
Lemma~\ref{lm:yuri-neretin-lemma}; the form $\rho$ and the set $B$ of Case~$1$; the integer points of the polytopes $R_C$ of Case~$2$).
Moreover, each step produces finitely many polyhedra, and the recursion has depth at most $2d$: each application of Case~$1$ lowers the
dimension of the polyhedron, each application of Case~$2$ lowers the dimension of the recession cones of the polyhedra which are further
decomposed, and neither dimension ever increases.
\end{remark}

 \noindent
 We are now able to finalize the proof of Theorem \ref{cor:MainDecomp}.
\medskip

\noindent
 {\bf Proof of Theorem \ref{cor:MainDecomp}.}
If $P$ is bounded, then $P\cap\Z^d$ is a finite set, which can be computed by enumeration, and \eqref{eq-main:cor:MainDecomp}
holds with $r_i=0$ for every $i$. Assume that $P$ is unbounded. By Proposition \ref{prop:MainDecomp}, one gets the partition
\begin{equation}\label{main-eq-Thm-6}
P \cap \Z^d \ = \ \bigsqcup _ {{\bf q} \in \Gamma} \left(({\bf q} + K _ {\q})\cap\Z^d\right),
\end{equation}
where $\Gamma\subseteq P\cap\Q^d$ is finite, the sets ${\bf q} + K_\q$, ${\bf q}\in\Gamma$, are pairwise disjoint, and each $K_\q$ is a
face of $K$, hence a pointed rational cone contained in $K$. By Proposition \ref{prop-refinement-cones}, applied to each of the finitely
many sets $({\bf q} + K_\q)\cap\Z^d$, every piece of \eqref{main-eq-Thm-6} is a finite disjoint union of simple sets
${\bf p} + {\bf v}_{1}^\oplus + \cdots + {\bf v}_{r}^\oplus$, where ${\bf p}\in\Z^d$ and ${\bf v}_1, \ldots, {\bf v}_r$ are linearly
independent vectors of $K_\q\cap\Z^d\subseteq K\cap\Z^d$. Collecting these decompositions, one obtains \eqref{eq-main:cor:MainDecomp},
the union being disjoint because the pieces of \eqref{main-eq-Thm-6} are pairwise disjoint. Finally, the decomposition is computable:
the family $\{{\bf q} + K_\q\}_{{\bf q}\in\Gamma}$ is computable by Remark \ref{rmk-Prop6}, and the decompositions of its members by
Proposition \ref{prop-refinement-cones}.
    \qed

\medskip

\noindent
The following example illustrates, on a polyhedron of dimension $3$, the construction of Proposition \ref{prop:MainDecomp} and the
proof of Theorem \ref{cor:MainDecomp}.

\begin{example}\label{ex:dim-3-case-2}
Let $P\subseteq \R^3$ be the polyhedron defined as
$$P= \{{\bf x} = (x, y, z) \in\R^3 : \ell_1({\bf x})=x+z\geq 1, \ \ell_2({\bf x})=y +z\geq 1,\ \ell_3({\bf x})=z-x\geq 0,\ \ell_4({\bf x})=z-2y\geq1\}.$$
It is easily checked that $P=M+K$, where $K$ is the recession cone of $P$ and $M=\mbox{conv}(\w_1, \w_2)$, with $\w_1=(0,0,1)$ and
$\w_2=(1,0,1)$. The extreme rays of $K$ are generated by
$${\bf a}=(-1,-1,1), \quad {\bf b}=(-2,1,2), \quad {\bf c}=(1,-1,1), \quad {\bf e}=(2,1,2),$$
so that $K$ is pointed and ${\rm dim}(K)=3$: we are therefore in Case $2$ of the proof of
Proposition \ref{prop:MainDecomp}. Note that $K$ is not simplicial (see Fig. 2, left, for its section by the plane $z=1$).

Since $\ell_3(\w_2)=0$, one has $P\subseteq K$ but $P\not\subseteq \mathring{K}$. According to Fact $1$, we first translate $P$ by the
vector ${\bf u}=(0,0,1)\in\Z^3$, obtaining
$$\widetilde P \ = \ \tau _ {\bf u}(P) \ = \ \{{\bf x}\in\R^3 \ : \ \ell_1({\bf x})\geq 2, \ \ell_2({\bf x})\geq 2, \
\ell_3({\bf x})\geq 1, \ \ell_4({\bf x})\geq 2\} \ \subseteq \ \mathring{K},$$
whose vertices are $\widetilde\w_1 = (0,0,2)$ and $\widetilde\w_2 = (1,0,2)$.

Let us now fix the point $\q = \widetilde\w_2 = (1,0,2)$ of $\widetilde M = {\rm conv}(\widetilde\w_1, \widetilde\w_2)$ and let us consider
the partition (\ref{prop:MainDecomp-lm-cone-eq}) of $K$ with respect to $\q$; here
$\gamma_1 = \ell_1(\q)=3$, $\gamma_2 = \ell_2(\q)=2$, $\gamma_3 = \ell_3(\q)=1$ and $\gamma_4 = \ell_4(\q)=2$. Since $\widetilde P$ is
defined by the inequalities $\ell_1\geq 2$, $\ell_2\geq 2 = \gamma_2$, $\ell_3\geq 1 = \gamma_3$ and $\ell_4\geq 2 = \gamma_4$, every
point of $\widetilde P$ carries the symbol $+$ in the positions $2, 3, 4$. Hence the only classes meeting $\widetilde P$ are those
defined by the words $(+,+,+,+)$ and $(-,+,+,+)$. Denoting by $C$ the latter, the partition (\ref{prop:MainDecomp-eq3}) of
$\widetilde P$ reduces to
$$\widetilde P \ = \ (\q + K) \ \sqcup \ (\widetilde P\cap C),$$
where $\q+K = \{{\bf x}\in\R^3 : \ell_1({\bf x})\geq 3, \ \ell_2({\bf x})\geq 2, \ \ell_3({\bf x})\geq 1, \ \ell_4({\bf x})\geq 2\}$ and
\begin{equation}\label{eq:last-example}
\widetilde P\cap C \ = \ \{{\bf x}\in\R^3 \ : \ 2\leq \ell_1({\bf x})< 3, \ \ell_2({\bf x})\geq 2, \
\ell_3({\bf x})\geq 1, \ \ell_4({\bf x})\geq 2\}
\end{equation}
is a semi-open polyhedron.

By Lemma \ref{lm:yuri-neretin-lemma}, the set (\ref{eq:last-example}) is replaced by a closed sub-polyhedron having the same integer
points. Since $\ell_1$ has integral coefficients, the construction of that lemma yields $\varepsilon_1 = 1$, that is the inequality
$\ell_1({\bf x})<3$ is replaced by $\ell_1({\bf x})\leq 2$; taking into account that $\ell_1({\bf x})\geq 2$, one gets
$$R_C \ = \ \{{\bf x}\in\R^3 \ : \ \ell_1({\bf x})= 2, \ \ell_2({\bf x})\geq 2, \ \ell_3({\bf x})\geq 1, \ \ell_4({\bf x})\geq 2\},
\quad\quad R_C\cap\Z^3 \ = \ \widetilde P\cap C\cap \Z^3.$$
A straightforward computation shows that the unique vertex of $R_C$ is $\widetilde\w_1 = (0,0,2)$ and that the recession cone of $R_C$ is
$$K_C \ = \ K \cap \{{\bf x}\in\R^3 \ : \ x+z=0\} \ = \ {\rm co}({\bf a}, {\bf b}),$$
a simplicial face of $K$ of dimension $2$, in accordance with Lemma \ref{constr-rec-con-Rv}. Thus $R_C = \widetilde\w_1 + K_C$ is
already a translate of a face of $K$, and the inductive hypothesis, applied to $R_C$, returns $R_C$ itself. Summing up,
$$\widetilde P\cap\Z^3 \ = \ \left((\q + K)\cap\Z^3\right) \ \sqcup \ \left((\widetilde\w_1 + K_C)\cap\Z^3\right),$$
which is the decomposition (\ref{prop:MainDecomp-eq0}) of Proposition \ref{prop:MainDecomp}, with
$\Gamma = \{(1,0,2), \ (0,0,2)\}$. Translating back by $-{\bf u}$, one finally obtains, for the original polyhedron $P$,
\begin{equation}\label{eq:ex-two-pieces}
P\cap\Z^3 \ = \ \left(\big((1,0,1) + K\big)\cap\Z^3\right) \ \sqcup \ \left(\big((0,0,1) + K_C\big)\cap\Z^3\right).
\end{equation}

Following the proof of Theorem \ref{cor:MainDecomp}, we now decompose the two pieces of \eqref{eq:ex-two-pieces} into simple sets.
The cone $K_C = {\rm co}({\bf a}, {\bf b})$ is simplicial, so that Proposition \ref{cor: transl-of-a-cone-simpl} applies directly. The
integer points of its fundamental parallelepiped are $(0,0,0)$, $\frac13({\bf a}+{\bf b}) = (-1,0,1)$ and
$\frac23({\bf a}+{\bf b}) = (-2,0,2)$, whence
$$\big((0,0,1) + K_C\big)\cap\Z^3 \ = \ \bigsqcup _ {{\bf p}\, \in\, \{(0,0,1),\, (-1,0,2),\, (-2,0,3)\}} \ \left({\bf p} + {\bf a}^\oplus + {\bf b}^\oplus\right).$$
The cone $K$ is not simplicial. Triangulating it into the simplicial cones ${\rm co}({\bf a}, {\bf b}, {\bf e})$ and
${\rm co}({\bf a}, {\bf e}, {\bf c})$, the proof of Proposition \ref{prop-refinement-cones} decomposes
$\big((1,0,1) + K\big)\cap\Z^3$ into $36$ pairwise disjoint simple sets: the apex $\{(1,0,1)\}$; one set for each of the four extreme
rays of $K$; $3$, $4$, $3$, $2$ and $1$ sets for the two-dimensional cones ${\rm co}({\bf a}, {\bf b})$, ${\rm co}({\bf b}, {\bf e})$,
${\rm co}({\bf e}, {\bf c})$, ${\rm co}({\bf c}, {\bf a})$ and ${\rm co}({\bf a}, {\bf e})$, respectively; $12$ and $6$ sets for the
three-dimensional cones ${\rm co}({\bf a}, {\bf b}, {\bf e})$ and ${\rm co}({\bf a}, {\bf e}, {\bf c})$, respectively (the apex
$(1,0,1)$ being integral, these are the numbers of the integer points of the fundamental parallelepipeds of the cones involved;
see Fig. 2).

\begin{center}
\begin{tikzpicture}[scale=1.6]
  \fill[gray!12] (-1,-1) -- (-1,0.5) -- (1,0.5) -- cycle;
  \fill[gray!28] (-1,-1) -- (1,0.5) -- (1,-1) -- cycle;
  \draw (-1,-1) -- (1,-1) -- (1,0.5) -- (-1,0.5);
  \draw[line width=1.6pt] (-1,-1) -- (-1,0.5);
  \draw[dashed] (-1,-1) -- (1,0.5);
  \foreach \p/\n/\pos in {{(-1,-1)}/a/below left,{(-1,0.5)}/b/above left,{(1,0.5)}/e/above right,{(1,-1)}/c/below right}
     {\fill \p circle (1.3pt); \node[\pos] at \p {${\bf \n}$};}
  \node[left] at (-1,-0.25) {\small $3$};
  \node[above] at (0,0.5) {\small $4$};
  \node[right] at (1,-0.25) {\small $3$};
  \node[below] at (0,-1) {\small $2$};
  \node[fill=white,inner sep=1pt] at (0,-0.25) {\small $1$};
  \node at (-0.45,0.12) {\small $12$};
  \node at (0.5,-0.62) {\small $6$};
  \node[left] at (-1.3,-0.25) {\small $K_C\,$};
\end{tikzpicture}
\hspace{1.2cm}
\begin{tikzpicture}[scale=0.62]
  \fill[gray!30] (-1,-2) rectangle (3,1);
  \draw (-2,-2) rectangle (3,1);
  \draw[line width=1.6pt] (-2,-2) -- (-2,1);
  \draw[densely dotted] (-1,-2) -- (-1,1);
  \fill (-2,-2) circle (2.4pt);
  \fill (-2,-1) circle (2.4pt);
  \fill (-2,0) circle (2.4pt);
  \fill (-2,1) circle (2.4pt);
  \fill (-1,-2) circle (2.4pt);
  \fill (-1,-1) circle (2.4pt);
  \fill (-1,0) circle (2.4pt);
  \fill (-1,1) circle (2.4pt);
  \fill (0,-2) circle (2.4pt);
  \fill (0,-1) circle (2.4pt);
  \fill (0,0) circle (2.4pt);
  \fill (0,1) circle (2.4pt);
  \fill (1,-2) circle (2.4pt);
  \fill (1,-1) circle (2.4pt);
  \fill (1,0) circle (2.4pt);
  \fill (1,1) circle (2.4pt);
  \fill (2,-2) circle (2.4pt);
  \fill (2,-1) circle (2.4pt);
  \fill (2,0) circle (2.4pt);
  \fill (2,1) circle (2.4pt);
  \fill (3,-2) circle (2.4pt);
  \fill (3,-1) circle (2.4pt);
  \fill (3,0) circle (2.4pt);
  \fill (3,1) circle (2.4pt);
  \node[below] at (-2,-2.1) {\scriptsize $x=-2$};
  \node[below] at (3,-2.1) {\scriptsize $x=3$};
  \node[left] at (-2.1,-2) {\scriptsize $y=-2$};
  \node[left] at (-2.1,1) {\scriptsize $y=1$};
  \node[above] at (1,1.05) {\small $(1,0,1)+K$};
  \node[above] at (-2,1.6) {\small $(0,0,1)+K_C$};
  \draw[->,thin] (-2,1.65) -- (-2,1.1);
\end{tikzpicture}

\smallskip
\parbox{0.92\textwidth}{\small {\bf Fig. 2.} Example \ref{ex:dim-3-case-2}. Left: the section of $K$ by the plane $z=1$, with the
rays ${\bf a}$, ${\bf b}$, ${\bf c}$, ${\bf e}$, the triangulation of $K$ into ${\rm co}({\bf a},{\bf b},{\bf e})$ and
${\rm co}({\bf a},{\bf e},{\bf c})$ (dashed diagonal), the face $K_C={\rm co}({\bf a},{\bf b})$ (thick), and the number of simple
sets contributed by the relative interior of each face of dimension $2$ and $3$. Right: the section of $P$ by the plane $z=3$, with its
integer points; the shaded rectangle is the section of $(1,0,1)+K$, the thick segment is the section of $(0,0,1)+K_C$, and the strip
between them contains no integer points.}
\end{center}

Altogether, $P\cap\Z^3$ is the disjoint union of $39$ simple sets, with integral apexes and with periods in
$\{{\bf a}, {\bf b}, {\bf c}, {\bf e}\}$, and, by Lemma \ref{lm:simple-dec}, the sum of the corresponding $39$ rational functions is
$\varphi ([P]) = \mathcal F([P])$. For instance, the three simple sets of the piece $(0,0,1) + K_C$ contribute
$$\frac{x_3 + x_1^{-1}x_3^{2} + x_1^{-2}x_3^{3}}{\left(1 - x_1^{-1}x_2^{-1}x_3\right)\left(1 - x_1^{-2}x_2\, x_3^{2}\right)}.$$
\end{example}

In Example \ref{ex:dim-3-case-2} only Case $2$ of the proof of Proposition \ref{prop:MainDecomp} occurs, and the polyhedron $R_C$
has dimension less than $d$. The next example illustrates Case $1$, both on its own and as it arises within Case $2$, when a
polyhedron $R_C$ has dimension $d$.

\begin{example}\label{ex:case-1}
(1) {\em Case $1$.} Let
$$P \ = \ \{(x,y)\in\R^2 \ : \ -1\leq x-2y\leq 2, \ \ x+2y\geq 1\}.$$
Then $P=M+K$, where $M$ is the segment with end points $(0,\frac12)$ and $(\frac32,-\frac14)$, and $K={\rm co}({\bf v})$, with
${\bf v}=(2,1)$. Since ${\rm dim}(K)=1<2={\rm dim}(P)$, we are in Case $1$, and we can take $\rho(x,y)=x-2y$, which vanishes on $K$. Then
$\rho(P)=\rho(M)=[-1,2]$ and $\rho(\Z^2)=\Z$, so that $B=\{-1,0,1,2\}$, and $P\cap\Z^2$ is the disjoint union of the sets
$P_\beta\cap\Z^2$, $\beta\in B$. Each slice $P_\beta=P\cap\{x-2y=\beta\}$ is the half-line $\q_\beta+K$, whose apex
$\q_\beta=\big(\frac{1+\beta}{2},\frac{1-\beta}{4}\big)$ is rational, and in general not integral; the inductive hypothesis (here, the
case of dimension $1$; see Fig. 3, left) returns $\Gamma_\beta=\{\q_\beta\}$ and $K_{\q_\beta}=K$. Hence
$$P\cap\Z^2 \ = \ \bigsqcup_{\beta\in B}\big((\q_\beta+K)\cap\Z^2\big), \qquad\quad
\Gamma \ = \ \big\{(0,\tfrac12), \ (\tfrac12,\tfrac14), \ (1,0), \ (\tfrac32,-\tfrac14)\big\}.$$
Since $K$ is simplicial, Proposition \ref{cor: transl-of-a-cone-simpl} applies directly: with $\Pi=\{\alpha{\bf v} : 0\leq\alpha<1\}$,
the set $(\q_\beta+\Pi)\cap\Z^2$ consists of the single point $(1,1)$, $(2,1)$, $(1,0)$, $(2,0)$ for $\beta=-1,0,1,2$, respectively.
Therefore
$$P\cap\Z^2 \ = \ \bigsqcup_{{\bf p}\,\in\,\{(1,1),\,(2,1),\,(1,0),\,(2,0)\}}\big({\bf p}+{\bf v}^\oplus\big),$$
and
$$\varphi([P]) \ = \ \mathcal F([P]) \ = \ \frac{x_1x_2+x_1^2x_2+x_1+x_1^2}{1-x_1^2x_2} \ = \ \frac{x_1(1+x_1)(1+x_2)}{1-x_1^2x_2}.$$

(2) {\em Case $1$ within Case $2$.} Let
$$P \ = \ \{(x,y)\in\R^2 \ : \ \ell_1(x,y)=x\geq 1, \ \ \ell_2(x,y)=y\geq 1, \ \ \ell_3(x,y)=x+y\geq 5\}.$$
Its recession cone is $K=\{x\geq 0, \ y\geq 0\}$, so that ${\rm dim}(K)=2$, and $P\subseteq\mathring{K}$, so that no translation is
needed. Take $\q=(4,1)$, a vertex of $P$, so that $\gamma_1=4$, $\gamma_2=1$ and $\gamma_3=5$. Every point of $P$ carries the symbol $+$
in the positions $2$ and $3$, so that only the chambers $(+,+,+)$, that is $\q+K$, and $C=(-,+,+)$ meet $P$, and
$P\cap C=\{1\leq x<4, \ y\geq 1, \ x+y\geq 5\}$. By Lemma \ref{lm:yuri-neretin-lemma}, the inequality $x<4$ is replaced by $x\leq 3$,
which gives
$$R_C \ = \ \{(x,y)\in\R^2 \ : \ 1\leq x\leq 3, \ \ y\geq 1, \ \ x+y\geq 5\}.$$
Its recession cone is $K_C={\rm co}({\bf e}_2)$, with ${\bf e}_2=(0,1)$, a face of $K$ of dimension $1$, in accordance with
Lemma \ref{constr-rec-con-Rv}. Unlike in Example \ref{ex:dim-3-case-2}, now ${\rm dim}(R_C)=2=d$, so that $R_C$ falls under Case $1$:
with $\rho(x,y)=x$ one gets $B=\{1,2,3\}$, and the three slices of $R_C$ are the half-lines $(\beta,5-\beta)+K_C$, $\beta=1,2,3$
(see Fig. 3, right).
Since all the cones involved are generated by vectors of a basis of $\Z^2$ and all the apexes are integral,
Proposition \ref{cor: transl-of-a-cone-simpl} gives one simple set for each piece, and
$$P\cap\Z^2 \ = \ \big((4,1)+{\bf e}_1^\oplus+{\bf e}_2^\oplus\big) \ \sqcup \ \bigsqcup_{\beta=1}^{3}\big((\beta,5-\beta)+{\bf e}_2^\oplus\big),$$
where ${\bf e}_1=(1,0)$, so that
$$\varphi([P]) \ = \ \mathcal F([P]) \ = \ \frac{x_1^4x_2}{(1-x_1)(1-x_2)} \ + \ \frac{x_1x_2^4+x_1^2x_2^3+x_1^3x_2^2}{1-x_2}.$$
\begin{center}
\begin{tikzpicture}[scale=0.62]
  \begin{scope}
  \clip (-0.8,-0.9) rectangle (7.6,4.3);
  \fill[gray!18] (0,0.5) -- (1.5,-0.25) -- (11.5,4.75) -- (10,5.5) -- cycle;
  \draw[thin] (0,0.5) -- ++(12,6);
  \draw[thin] (0.5,0.25) -- ++(12,6);
  \draw[thin] (1,0) -- ++(12,6);
  \draw[thin] (1.5,-0.25) -- ++(12,6);
  \end{scope}
  \draw[->] (-0.8,0) -- (7.9,0) node[right] {\small $x$};
  \draw[->] (0,-0.9) -- (0,4.6) node[above] {\small $y$};
  \draw[line width=1.6pt] (0,0.5) -- (1.5,-0.25);
  \fill (3,1) circle (2.3pt);
  \fill (3,2) circle (2.3pt);
  \fill (4,1) circle (2.3pt);
  \fill (4,2) circle (2.3pt);
  \fill (5,2) circle (2.3pt);
  \fill (5,3) circle (2.3pt);
  \fill (6,2) circle (2.3pt);
  \fill (6,3) circle (2.3pt);
  \fill (7,3) circle (2.3pt);
  \fill (7,4) circle (2.3pt);
  \draw[fill=white] (0,0.5) circle (2.8pt);
  \draw[fill=white] (0.5,0.25) circle (2.8pt);
  \draw[fill=white] (1,0) circle (2.8pt);
  \draw[fill=white] (1.5,-0.25) circle (2.8pt);
  \fill (0.87,0.87) rectangle (1.13,1.13);
  \fill (1.87,0.87) rectangle (2.13,1.13);
  \fill (0.87,-0.13) rectangle (1.13,0.13);
  \fill (1.87,-0.13) rectangle (2.13,0.13);
  \node[left] at (6.6,3.95) {\scriptsize $\beta=-1$};
  \node[right] at (7.6,2.8) {\scriptsize $\beta=2$};
  \node at (0.55,-0.5) {\small $M$};
\end{tikzpicture}
\hspace{0.8cm}
\begin{tikzpicture}[scale=0.52]
  \fill[gray!12] (1,4) -- (4,1) -- (7.5,1) -- (7.5,7.5) -- (1,7.5) -- cycle;
  \fill[gray!35] (4,1) rectangle (7.5,7.5);
  \fill[gray!22] (1,4) -- (3,2) -- (3,7.5) -- (1,7.5) -- cycle;
  \draw[->] (0,0) -- (8,0) node[right] {\small $x$};
  \draw[->] (0,0) -- (0,8) node[above] {\small $y$};
  \draw (1,7.5) -- (1,4) -- (4,1) -- (7.5,1);
  \draw[dashed] (4,0) -- (4,7.7);
  \foreach \b in {1,2,3} {\draw[line width=1.2pt] (\b,5-\b) -- (\b,7.5);}
  \fill (1,5) circle (2.2pt);
  \fill (1,6) circle (2.2pt);
  \fill (1,7) circle (2.2pt);
  \fill (2,4) circle (2.2pt);
  \fill (2,5) circle (2.2pt);
  \fill (2,6) circle (2.2pt);
  \fill (2,7) circle (2.2pt);
  \fill (3,3) circle (2.2pt);
  \fill (3,4) circle (2.2pt);
  \fill (3,5) circle (2.2pt);
  \fill (3,6) circle (2.2pt);
  \fill (3,7) circle (2.2pt);
  \fill (4,2) circle (2.2pt);
  \fill (4,3) circle (2.2pt);
  \fill (4,4) circle (2.2pt);
  \fill (4,5) circle (2.2pt);
  \fill (4,6) circle (2.2pt);
  \fill (4,7) circle (2.2pt);
  \fill (5,1) circle (2.2pt);
  \fill (5,2) circle (2.2pt);
  \fill (5,3) circle (2.2pt);
  \fill (5,4) circle (2.2pt);
  \fill (5,5) circle (2.2pt);
  \fill (5,6) circle (2.2pt);
  \fill (5,7) circle (2.2pt);
  \fill (6,1) circle (2.2pt);
  \fill (6,2) circle (2.2pt);
  \fill (6,3) circle (2.2pt);
  \fill (6,4) circle (2.2pt);
  \fill (6,5) circle (2.2pt);
  \fill (6,6) circle (2.2pt);
  \fill (6,7) circle (2.2pt);
  \fill (7,1) circle (2.2pt);
  \fill (7,2) circle (2.2pt);
  \fill (7,3) circle (2.2pt);
  \fill (7,4) circle (2.2pt);
  \fill (7,5) circle (2.2pt);
  \fill (7,6) circle (2.2pt);
  \fill (7,7) circle (2.2pt);
  \fill (3.85,0.85) rectangle (4.15,1.15);
  \fill (0.85,3.85) rectangle (1.15,4.15);
  \fill (1.85,2.85) rectangle (2.15,3.15);
  \fill (2.85,1.85) rectangle (3.15,2.15);
  \node[below right] at (4,1) {\small $\q$};
  \node[fill=gray!35,inner sep=1pt] at (5.5,3.5) {\small $\q+K$};
  \node[above] at (2,7.5) {\small $R_C$};
  \node[below] at (4,0) {\small $x=4$};
\end{tikzpicture}

\smallskip
\parbox{0.92\textwidth}{\small {\bf Fig. 3.} Example \ref{ex:case-1}. Left, part (1): the polyhedron $P$ (shaded), the segment $M$,
the four slices $\q_\beta+K$, $\beta=-1,0,1,2$, with their rational apexes $\q_\beta$ (white dots), and the integral apexes of the
four simple sets (squares). Right, part (2): the polyhedron $P$, the piece $\q+K$ (dark), the line $x=4$ separating the chambers,
the closed polyhedron $R_C$ (medium) with its three slices (thick), and the apexes of the four simple sets (squares).}
\end{center}
\end{example}

          \section{A theorem of Brion}\label{sec:brion}
           As pointed out in the Introduction,  an alternative proof of Theorem \ref{th:MAIN-B} can be obtained by using  a result named {\em Brion's identity}. 
Broadly speaking, this identity expresses the indicator function of a polyhedron, up to polyhedra with lines, through the tangent
cones at its vertices; indicator functions are added and subtracted there, according to the Inclusion-Exclusion principle.
More precisely, the following result holds.
\begin{theorem}{}\label{thm:Brion}
Let $P$ be a rational polyhedron. Then one has
       \begin{equation}\label{eq:Brion}
[P] \ = \  \sum _ {{\bf v} \in Vert(P)}\ [{\bf v} + K_{\bf v}] \ + \  {\cal G},
\end{equation}
where $Vert(P)$ is the set of vertices of $P$, $K_{\bf v}$ is the tangent cone of $P$ at $\bf v$ (${\bf v} \in Vert(P)$), that is, if $P$ is described by Eq. (\ref{lm9:eq0-polyhedron}), the cone
$K_{\bf v} = \{{\bf x}\in\R^d : \ell _i({\bf x})\geq 0 \ \mbox{for every $i\in I$ such that} \ \ell _i({\bf v}) = \gamma _i\}$, and $\cal G$ is a linear combination of indicator functions of rational polyhedra with lines.

 \end{theorem}

For the original proof, based on toric varieties, see \cite{Brion}; for elementary proofs, see
\cite[Corollary 4.6]{Ba-02} and \cite[Theorem 6.4]{Bar-08}.

Brion's identity yields the following alternative proof of Theorem \ref{th:MAIN-B}.

     \begin{proof}[Second proof of Theorem \ref{th:MAIN-B}.]
Let $P$ be a rational polyhedron. By Theorem \ref{th:BR} and Theorem \ref{th:KPL}, $\varphi$ and $\mathcal F$ are valuations which
vanish on the indicator functions of the polyhedra containing lines. Hence, applying them to Eq. (\ref{eq:Brion}), one gets
$$\varphi ([P]) \ = \ \sum _ {{\bf v} \in Vert(P)}\ \varphi ([{\bf v} + K_{\bf v}]), \quad\quad\quad
\mathcal F ([P]) \ = \ \sum _ {{\bf v} \in Vert(P)}\ \mathcal F ([{\bf v} + K_{\bf v}]).$$
Every vertex ${\bf v}$ of $P$ is a rational point (in general, not an integral one) and its tangent cone $K_{\bf v}$ is a pointed
rational cone, so that $\varphi ([{\bf v} + K_{\bf v}]) = \mathcal F ([{\bf v} + K_{\bf v}])$ by
Proposition \ref{prop:BR-affine-cones}. Therefore $\varphi ([P]) = \mathcal F ([P])$. Note that this argument uses only the case of
cones of Section \ref{sec:cones}, and not the decomposition of Theorem \ref{cor:MainDecomp}.
            \qed  \end{proof}

 \medskip

\noindent{\bf\large Acknowledgments}
\vskip5pt
{We thank Prof. S. Brlek and Ch. Reutenauer for their comments
on an earlier version of the manuscript and for insightful discussions on the topics of
this paper.}
\vskip5pt
\noindent{\bf\large AI-generated content disclosure}

In this work, we used generative AI tools for double-checking our results against existing literature. We take responsibility for the final content of this work.

\vskip5pt
\noindent
\noindent{\bf\large Data Availability}
\vskip5pt
Data sharing not applicable to this article  since no datasets were generated or analyzed during the current study.

\end{document}